\documentclass{svjour2}                    
\smartqed  
\usepackage{dcolumn}
\usepackage{bm}
\usepackage{graphicx}
\usepackage{color}
\usepackage{subfigure}
\usepackage{hyperref}
\usepackage{latexsym}
\usepackage{amsfonts}
\usepackage{cite}

\DeclareGraphicsExtensions{.jpg,.pdf, .mps, .png, .eps, .ps, .EPS,.gif}

\begin{document}

\def\be{\begin{equation}}
\def\ee{\end{equation}}

\def\bc{\begin{center}}
\def\ec{\end{center}}
\def\bea{\begin{eqnarray}}
\def\eea{\end{eqnarray}}
\newcommand{\avg}[1]{\langle{#1}\rangle}
\newcommand{\Avg}[1]{\left\langle{#1}\right\rangle}

\def\ie{\textit{i.e.}}
\def\etal{\textit{et al.}}
\def\m{\vec{m}}
\def\G{\mathcal{G}}

\newcommand{\gin}[1]{{\bf\color{blue}#1}}
\newcommand{\daan}[1]{{\bf\color{green}#1}}

\title{Network Geometry and Complexity}


\author{Daan Mulder       \and
        Ginestra Bianconi 
}


\institute{D. Mulder \at
			School of Mathematical Sciences\\
              Queen Mary University of London \\
           \and
           G. Bianconi \at
           School of Mathematical Sciences\\
              Queen Mary University of London \\
               \email{ginestra.bianconi@gmail.com}           
             }

\date{Received: date / Accepted: date}

\maketitle

\begin{abstract}
Higher order networks are able to characterize data as different as functional brain networks, protein interaction networks and social networks beyond the framework of pairwise interactions.
Most notably higher order networks include simplicial complexes formed not only by nodes and links but also by   triangles, tetrahedra, etc. More in general, higher-order networks can be cell-complexes formed by gluing convex polytopes along their faces.  Interestingly, higher order networks have a natural geometric interpretation and therefore constitute a natural way to explore the discrete network geometry of complex networks. Here we investigate the rich interplay between emergent network geometry of higher order networks and their complexity in the  framework of a non-equilibrium model called Network Geometry with Flavor. This model, originally proposed for capturing the evolution of simplicial complexes, is here extended to cell-complexes formed by subsequently gluing  different copies of an arbitrary regular polytope. We  reveal the interplay between complexity and geometry  of the higher order networks generated by the model by studying  the emergent community structure and  the degree distribution as a function of the  regular polytope forming its building blocks. Additionally, we discuss the underlying hyperbolic nature of the emergent geometry and we relate the spectral dimension of the higher-order network to the dimension and nature of its building blocks.
\end{abstract}

\section{Introduction}

Network Science \cite{BA,SW,Doro_book,Newman_book,Laszlo_book} has allowed an incredible progress in the understanding of the underlying architecture of complex systems and is having profound implications for different fields ranging from brain research \cite{Bassett_Sporns} and network medicine \cite{NetMedicine} to global infrastructures \cite{Havlin}.

It is widely believed \cite{perspective} that in  order to advance further in our understanding of complex systems it is important to consider generalized networks structures. These include both multilayer networks formed by several interacting networks \cite{PhysRep,Kivela} and higher order networks which allow going beyond the framework of pairwise interactions \cite{Bassett,Emergent,NGF,Hyperbolic,CQNM,PRE,Courtney2,Doro_manifolds,Zlatic,Courtney,Krioukov,Nechaev,Kahle,Costa1,Costa2}.

Higher order networks can be essential when analyzing brain networks \cite{Bassett,Torre,Vaccarino1,Vaccarino2,Giusti}, protein interaction  networks \cite{proteins} or social networks \cite{Barrat1,Barrat2}. For instance in brain functional networks, it is important to distinguish between brain regions that interact as a pair, or as a part of a larger complex, yielding their simultaneous co-activation \cite{Bassett}.  
Similarly, protein interaction networks map the relations between protein complexes of the cell, which are formed by several connected proteins that are able to perform a specific biological function \cite{proteins}. In social networks simplicial complexes arise in different contexts \cite{Social1,Social2,Barrat1,Barrat2}, as for instance in face-to-face interacting networks constituted by small  groups that form and dissolve in time, usually including  more than two people \cite{Barrat1,Barrat2}.

In many cases  the building blocks of a higher order network structures are   $d$-dimensional simplices such as triangles, tetrahedra etc., i.e. a set of $(d+1)$ nodes in which each node is interacting with all the others. In this case higher order networks are called simplicial complexes. However, there are some occasions in which it is important to consider higher order networks formed by building blocks that are less densely connected than simplices, i.e cell-complexes formed by gluing convex polytopes. Cell-complexes are of fundamental importance for characterizing self-assembled nanostructures \cite{Tadic} or  granular materials \cite{Granular}. However examples where cell-complexes are relevant also for  interdisciplinary applications are not lacking.  For instance, a protein complex is formed by a set of connected proteins, but not all proteins necessarily bind to every other protein in the complex. Also in face-to-face interactions, a social gathering of people might be organized into small groups, where each group  can include people that do not know each other directly. These considerations explain the need to extend the present modelling framework from simplicial complexes to general cell-complexes  formed by  regular polytopes such  as cubes, octahedra etc. 

 Modelling frameworks for simplicial complexes \cite{Bassett,Emergent,NGF,Hyperbolic,CQNM,PRE,Courtney2,Doro_manifolds,Zlatic,Courtney,Krioukov,Nechaev,Kahle,Costa1,Costa2} include both equilibrium static models that can be used as null models \cite{Zlatic,Courtney,Krioukov,Nechaev,Kahle,Costa1,Costa2,Patania} and non-equilibrium growing models describing their temporal evolution \cite{Emergent,NGF,Hyperbolic,PRE,CQNM,Courtney2,Doro_manifolds}. 
However, modelling of cell-complexes has been mostly neglected by the network science community.

Characterizing non-equilibrium growth models of cell-complexes allows us to  investigate the relation between the local geometrical structure of the higher order networks and their global  properties, revealing the nature of their emergent geometry and their complexity.

Interestingly, simplices and more in general convex polytopes  have a natural geometrical interpretation and are therefore essential for investigating  network geometry \cite{Emergent,Hyperbolic}.  As such simplicial complexes are widely adopted in quantum gravity for investigating the geometry of space-time \cite{Loll,Oriti,Tensor}. Network geometry is also a topic of increasing interest for network scientists which aim at gaining further understanding of  discrete network structures using geometry. This field is expected to provide deep insights and solid mathematical foundation to the characterization of the community structure of networks  \cite{Santo,Iacovacci,Redner}, contribute in inference problems \cite{Cannistraci} and shed new light onto the relation between structure (and in specific network geometry) and dynamics \cite{Torre,Ana}.

The recent interest in network geometry  is reflected in the vibrant research activity which aims at defining   the curvature of networks  and at extracting geometrical information from network data using these definitions \cite{Yau1,Yau2,Gromov,Jost1,Jost2,Jost3,Curvature_loll}. In a variety of cases \cite{Aste,Kleinberg,Boguna_Internet,Boguna_metabolic} it has been claimed that actually the underlying network geometry of complex networks is hyperbolic \cite{Nechaev2}. This hidden hyperbolic geometry is believed to be very beneficial for routing algorithms  and navigability \cite{Kleinberg,Boguna_Internet,Boguna_navigability}.
While several equilibrium and non-equilibrium models imposing an underlying hyperbolic network geometry have been widely studied and applied to real networks \cite{Boguna_hyperbolic,Boguna_growing}, recently a significant progress has been made in characterizing emergent hyperbolic network structures \cite{Hyperbolic}.
In particular it has been found that  Network Geometry with Flavor \cite{NGF} is a comprehensive theoretical framework that provides a main avenue to explore emergent hyperbolic geometry \cite{Hyperbolic}. This model uses a non-equilibrium evolution of simplicial complexes that is purely combinatorial, i.e. it makes no assumptions on the underlying geometry.  The  hyperbolic network geometry of the resulting structure is not a priori assumed but instead it is an emergent property of the network evolution. 
  
The theoretical framework of Network Geometry with Flavor shows that non-equilibrium growth dictated by purely combinatorial and probabilistic rules is able to generate an hyperbolic network geometry, and at the same time  determines a comprehensive theoretical framework able to generate very different network structures including chains, manifolds, and networks growing with preferential attachment. Most notably this model  includes as limiting cases models that until now have been  considered to be completely independent such as  the Barab\'asi-Albert model \cite{BA} and the random Apollonian network \cite{apollonian,apollonian_ising,apollonian2,Apollonian_group}.

In this paper we extend the Network Geometry with Flavor originally formulated for simplicial complexes to cell-complexes  formed by  any type of regular polytopes.
In particular we will focus on Network Geometry with Flavor $s\in \{-1,0,1\}$ built by subsequently gluing different copies of a  regular polytope along its faces. Note that in this paper we conside cell-complexes formed by an arbitrary  regular polytope  but any cell-complex is {\it pure}, i.e. it has only one type of regular polytope forming its building blocks. 

Although cell-complexes can be in several occasions a realistic representation of network data, is not our intention to propose a very realistic model of cell-complexes. Rather our goal is on one side to propose a very simple theoretical model for  emergent geometry  and on the other side to investigate the interplay between its geometry and its complexity. 

The network geometry is investigated by characterizing the Hausdorff, the spectral \cite{Burioni,Spectral,Benedetti,Thomas} and the cell-complex' topological dimension, together with the  "holographic" nature of the model.  The complexity of the resulting  network structures is studied by deriving under which conditions the resulting networks are scale-free and display a non-trivial  emergent community structure.

Finally, Network Geometry with Flavor can be considered as the natural extension of the very widely studied framework of non-equilibrium growing complex networks models (with and without preferetial attachment) to characterize network geometry in any dimension. In this respect many non-trivial results are obtained. For instance, we show that when working with simplices scale-free networks can emerge from a dynamical rule that does not contain an explicit preferential attachment mechanism. Additionally, we reveal that even when preferential attachment of the regular polytopes is  present, the Network Geometry with Flavor might result in a homogeneous network structure in which  the second moment of the degree distribution is finite in the large network limit.

\section{Simplicial Complexes and Higher Order Networks}

Simplicial complexes provide the main example of higher order networks where interactions are not only pairwise, but can include more than two nodes.
Simplicial complexes are formed by simplices glued along their faces.
A simplex of dimension $d$ is  a set of $d+1$ nodes and describes the many-body interaction between these nodes.
A simplex admits a natural geometrical interpretation. For instance a  simplex of dimension $d=0,1,2,3$, can be identified with a node, a link, a triangle and a tetrahedron respectively.
A $\delta$-dimensional face of a  simplex $\alpha$ of dimension $d>\delta$ is a simplex $\hat{\alpha}$ formed by a subset of $\delta+1$ of the nodes of $\alpha$. 
A simplicial complex $\mathcal{K}$ of dimension $d$ is formed by a set of  simplices of dimension $\delta\leq d$ glued along their faces.   Additionally, this set must be closed under the operation of taking faces of any simplex.
Therefore, in mathematical terms it must satisfy two conditions: \begin{itemize}
\item[a)] the intersection $\alpha \cap \tilde{\alpha}$ of two simplices $\alpha\in {\mathcal K}$ and $\tilde{\alpha}\in {\mathcal K}$ belonging to the simplicial complex  is a simplex of the simplicial complex, i.e. $\alpha \cap \tilde{\alpha}\in {\mathcal K}$; 
\item[b)] if the simplex $\alpha$ belongs to the simplicial complex, i.e. $\alpha\in {\mathcal K}$, then every simplex $\hat{\alpha}$ which is a face of $\alpha$ (i.e. $\hat{\alpha}\subset \alpha$)  must also belong to the simplicial complex, i.e. $\hat{\alpha}\in {\mathcal K}$. 
\end{itemize}
Among simplicial complexes we distinguish {\em pure} $d$-dimensional simplicial complexes which are formed exclusively by $d$-dimensional simplices and their faces. 

Here we consider not only simplicial complexes, but we treat also cell-complexes, which differ from simplicial complexes because they are formed by subsequently gluing convex polytopes along their faces.  In particular we will focus on cell-complexes ${\mathcal Q}$ formed by identical $d$-dimensional regular polytopes glued  along their $(d-1)$-faces,  called here {\it pure cell-complexes}. A pure cell-complex   reduce to  pure $d$-dimensional simplicial complex if the regular polytope that constitute its building blocks is a $d$-dimensional simplex \cite{Note2}. 

A regular polytope of dimension $d$ is a  maximally symmetric $d$-dimensional polytope having $F$  identical $(d-1)$-dimensional  faces and $M$ nodes. Each   node of a regular polytope has degree $v$ and it is  incident to the same  number $f$ of $(d-1)$-dimensional faces. Each $(d-1)$-face includes $m$ nodes.
A $d$-dimensional simplex is a regular polytope.  However, the number of regular polytopes in dimension $d\geq 2$ is larger than one.  In  Table $\ref{table:polytopes}$ we report  the complete list of regular polytopes and their properties.

\begin{figure}
$\begin{array}{ccccc}
  \includegraphics[width=0.2\columnwidth]{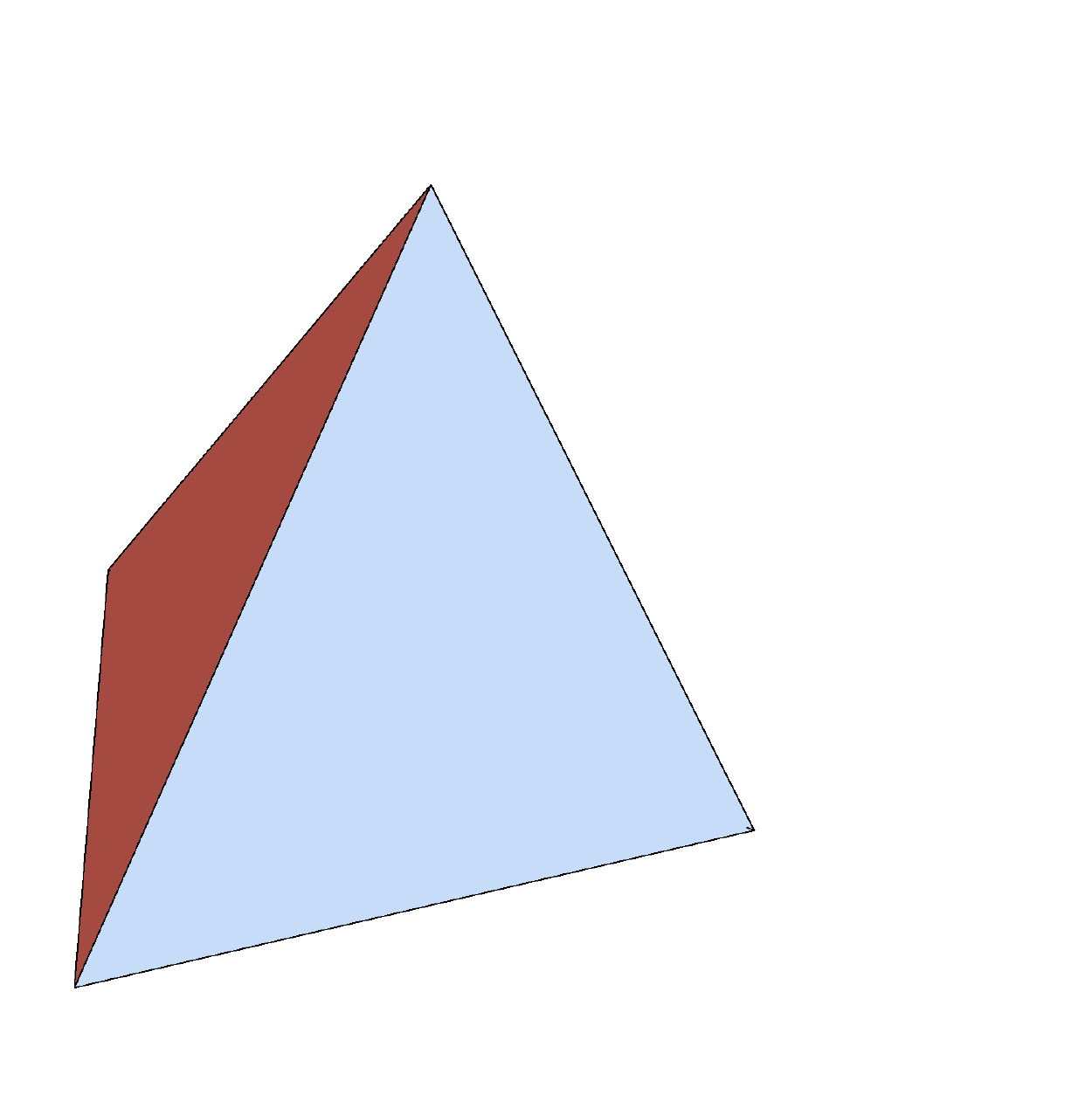}&
  \includegraphics[width=0.2\columnwidth]{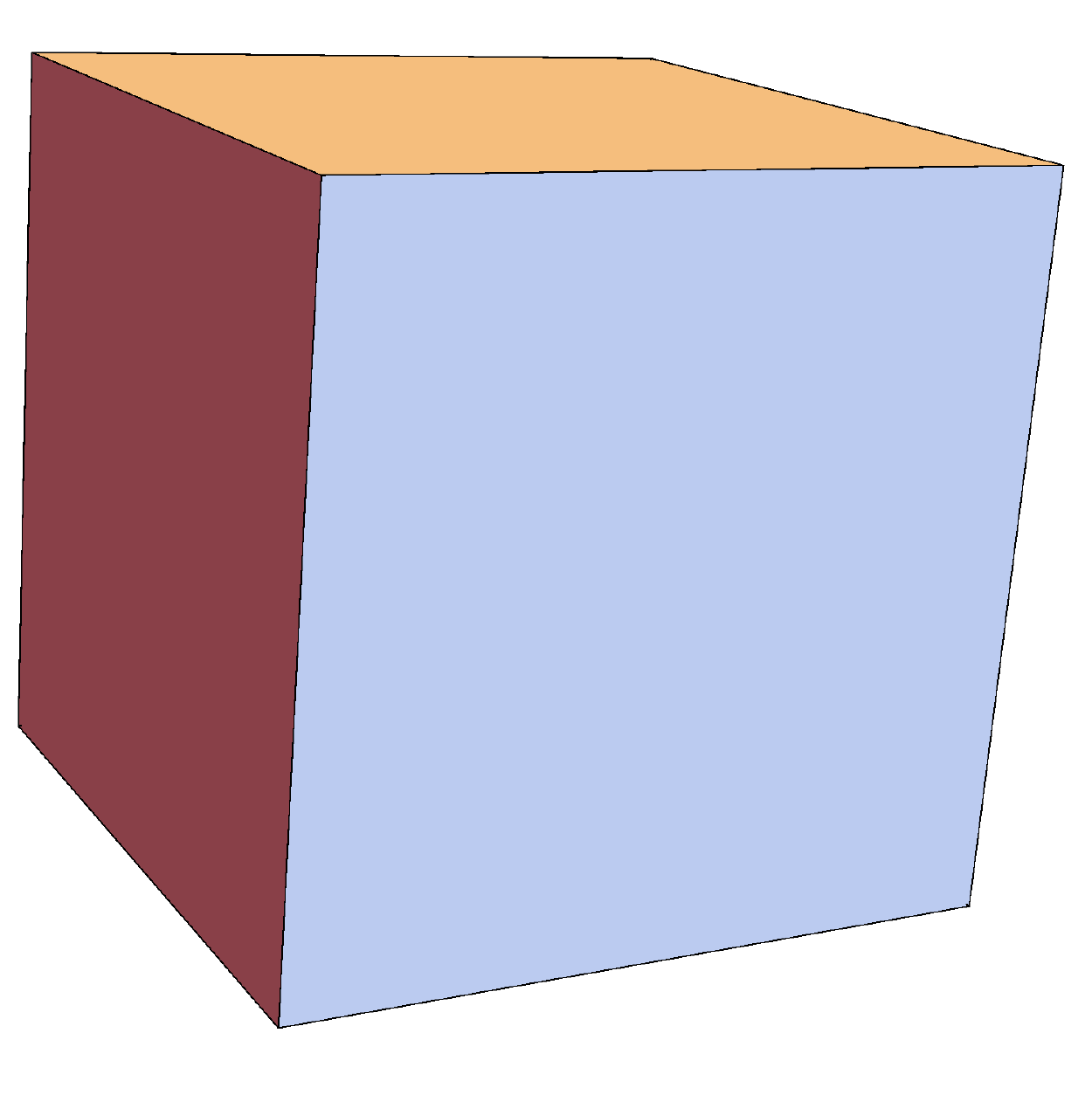}&
  \includegraphics[width=0.2\columnwidth]{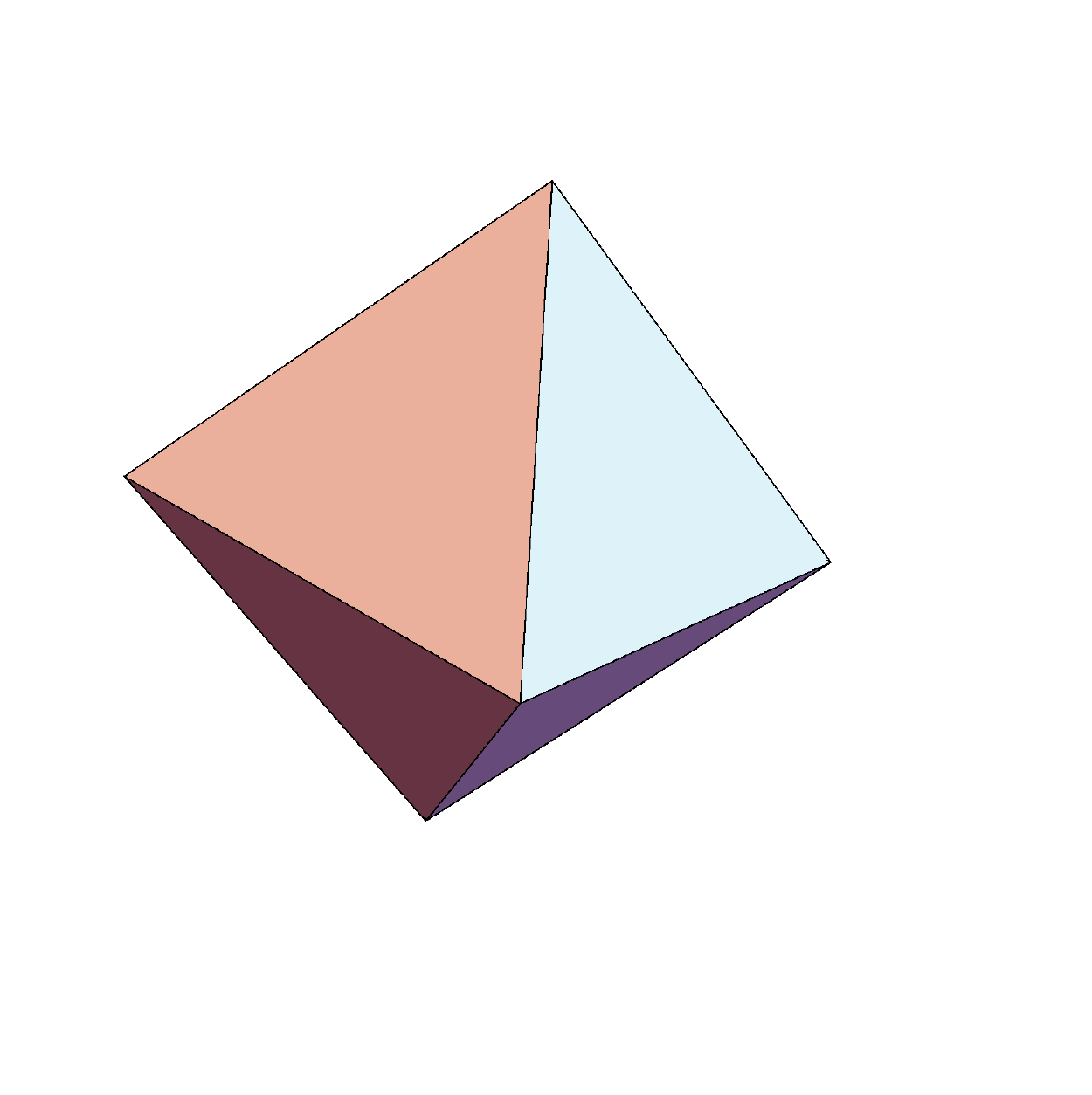}&
  \includegraphics[width=0.2\columnwidth]{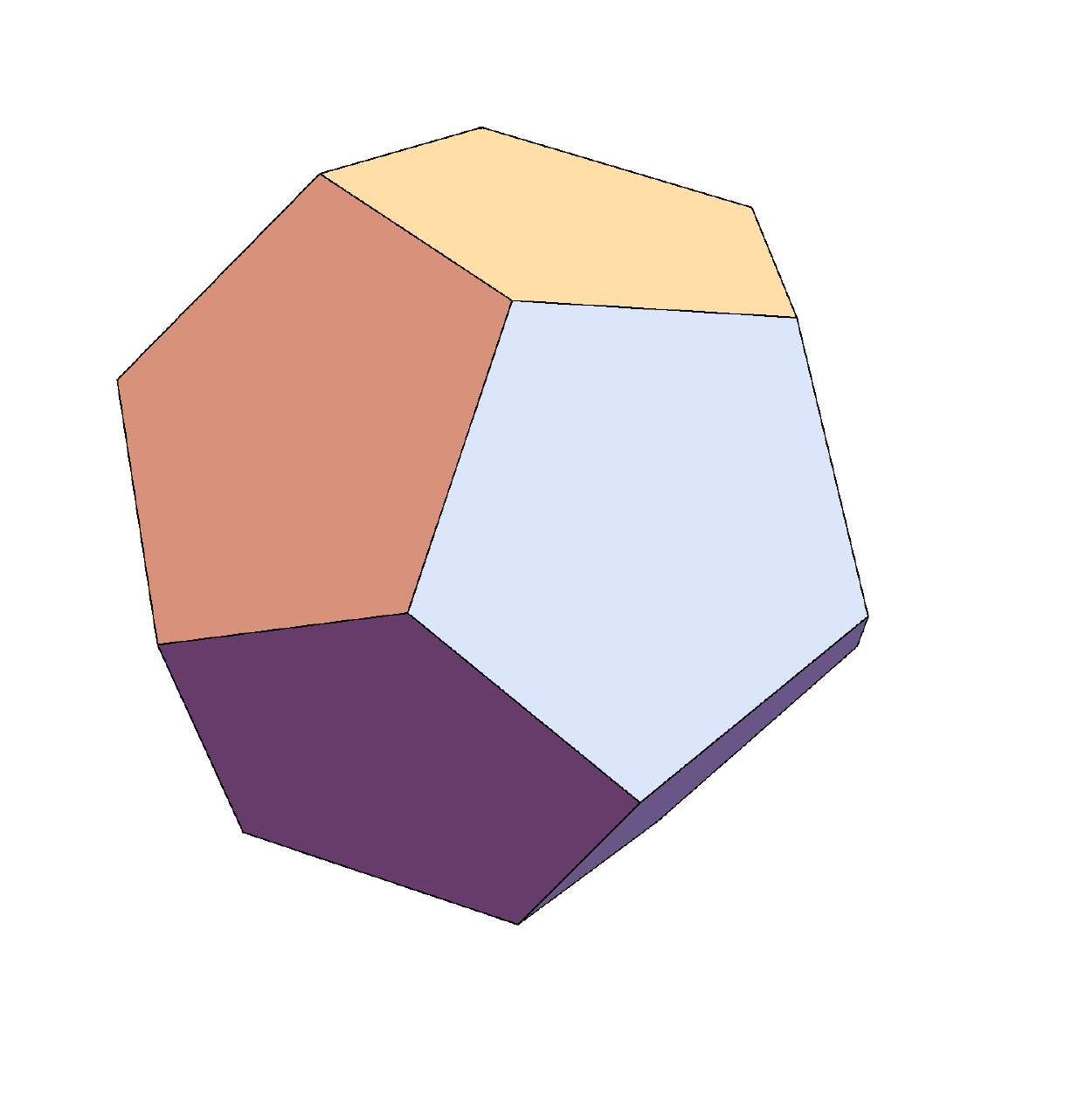}&
  \includegraphics[width=0.2\columnwidth]{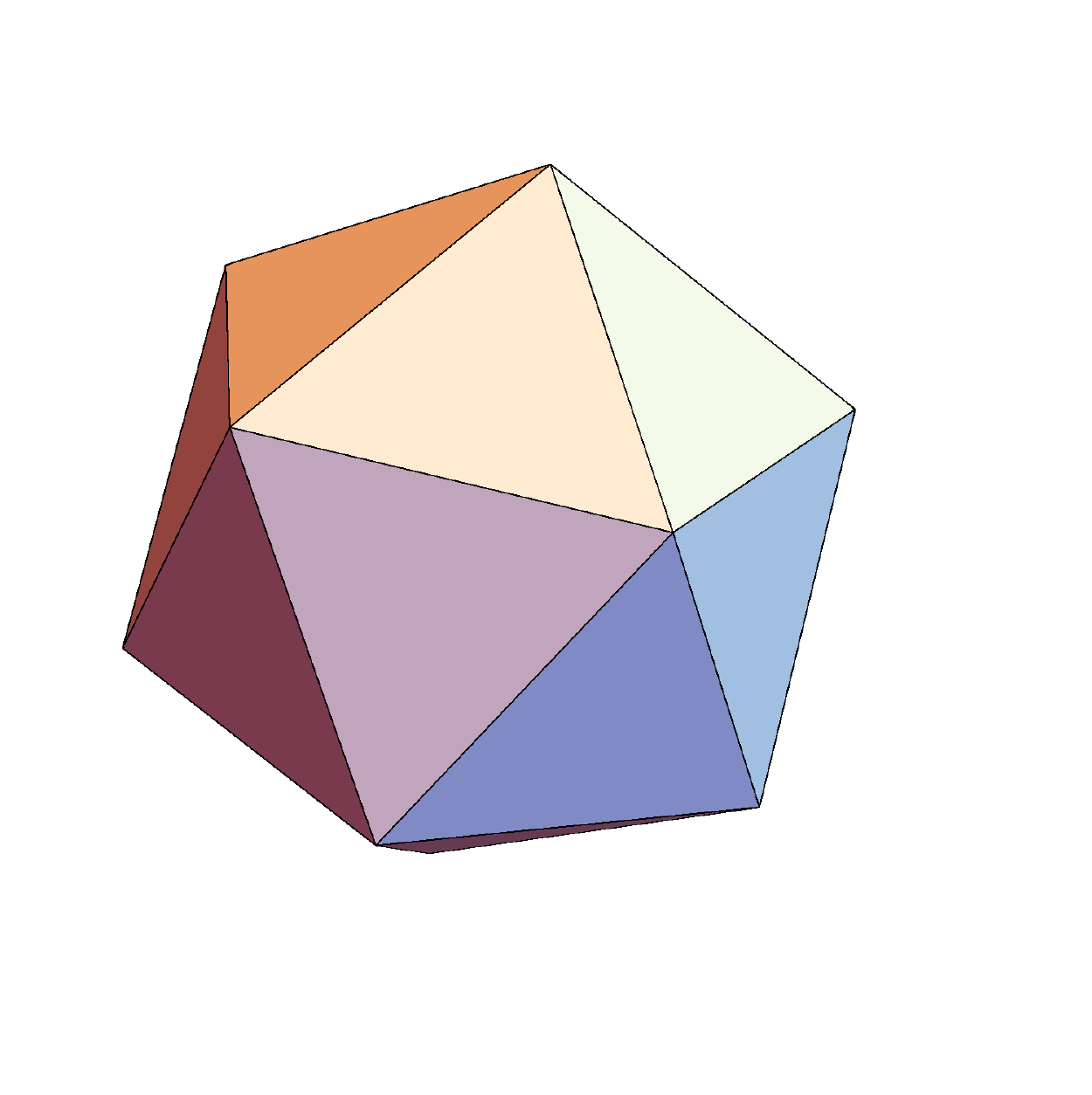}
  \nonumber \\
    \mbox{\bf (a)}&  \mbox{\bf (b)}&  \mbox{\bf (c)}&  \mbox{\bf (d)}&  \mbox{\bf (e)}
  \end{array}$
\caption{All the regular polytopes in $d=3$ (Platonic solids): (a) tetrahedron, (b) cube, (c) octahedron, (d) dodecahedron, (e) icosahedron.}
\label{fig:Platonic}       
\end{figure}

\begin{figure}
$\begin{array}{ccccc}
  \includegraphics[width=0.2\columnwidth]{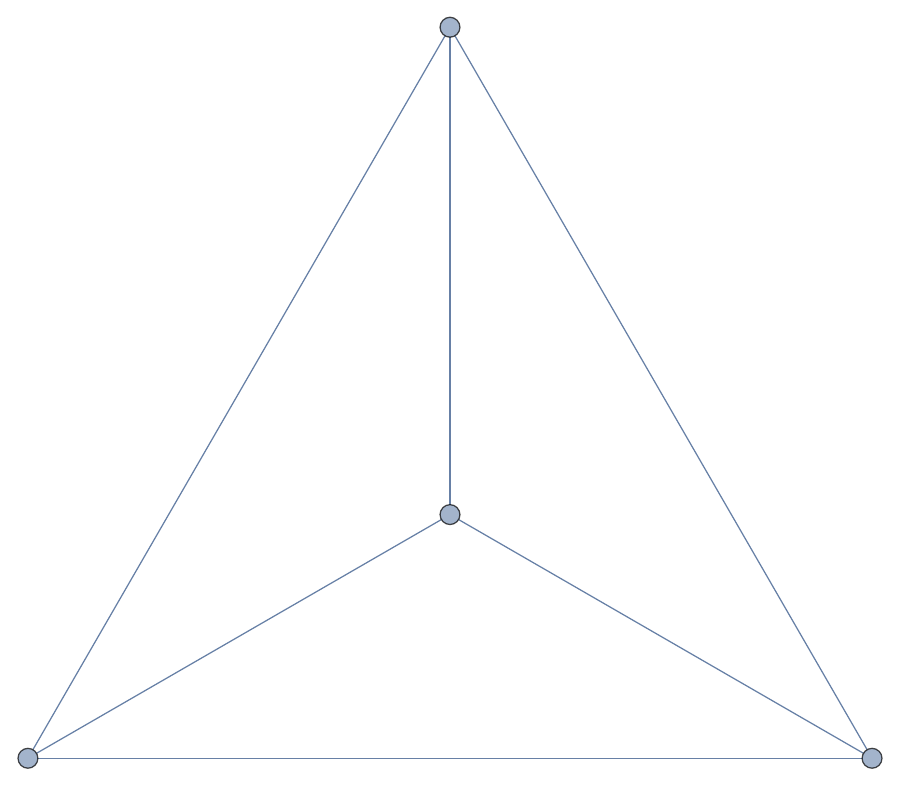}&
  \includegraphics[width=0.2\columnwidth]{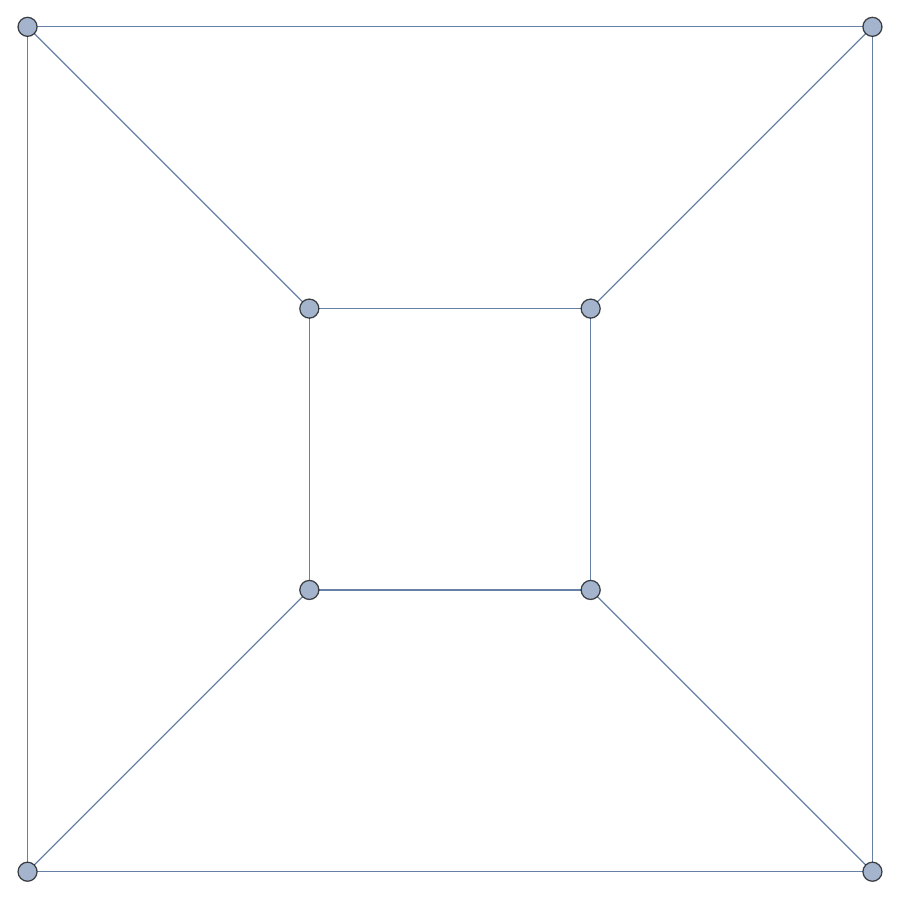}&
  \includegraphics[width=0.2\columnwidth]{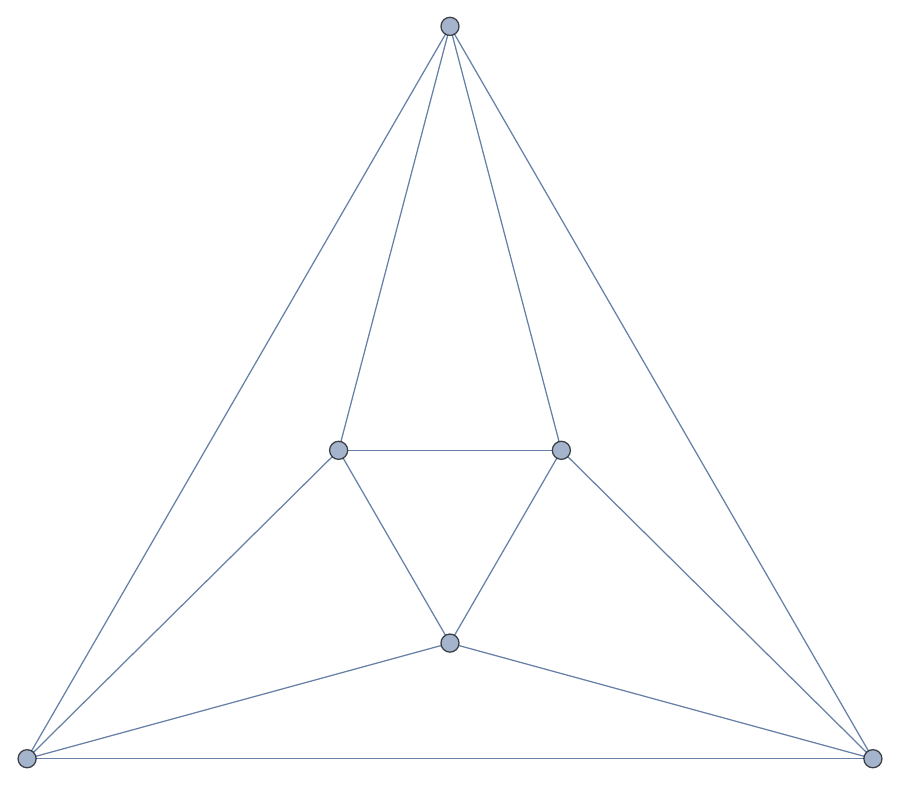}&
  \includegraphics[width=0.2\columnwidth]{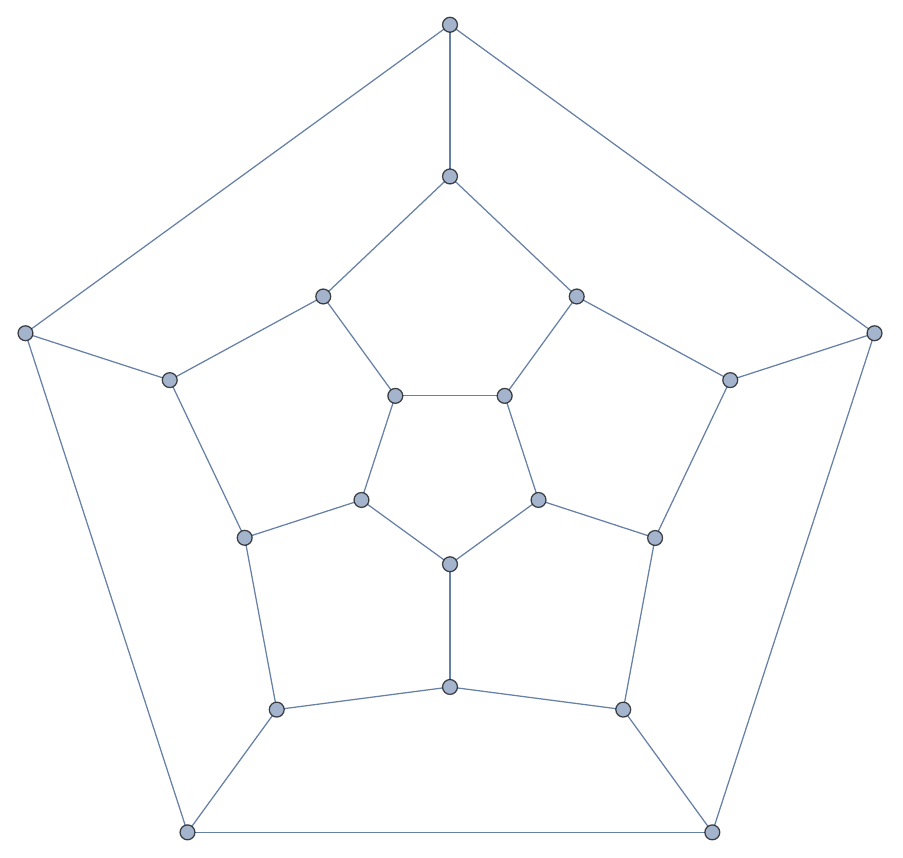}&
  \includegraphics[width=0.2\columnwidth]{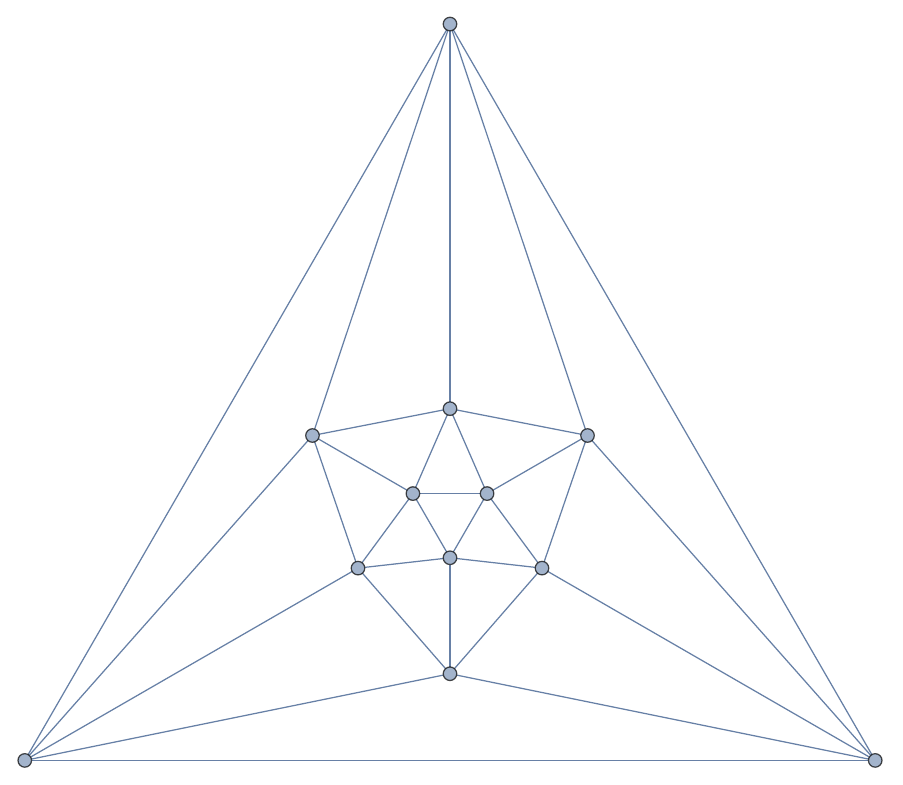}
  \nonumber \\
  \mbox{\bf (a)}&  \mbox{\bf (b)}&  \mbox{\bf (c)}&  \mbox{\bf (d)}&  \mbox{\bf (e)}
  \end{array}$
\caption{The planar graphs that describe the skeleton of the regular polytopes in $d=3$ (Platonic solids): (a) tetrahedron, (b) cube, (c) octahedron, (d) dodecahedron, (e) icosahedron.}
\label{fig:planar}      
\end{figure}

\begin{itemize}
\item[(1)]{\bf Dimension  $d=1$}\\
This is the trivial case in which the regular polytope is just a single link.\\
\item[(2)]{\bf Dimension  $d=2$}\\
In dimension $d=2$ the  regular polytopes are the regular polygons, i.e. triangles, squares, pentagons, hexagons etc.\\ 
\item[(3)]{\bf Dimension  $d=3$}\\
In dimension $d=3$ the regular polytopes are the Platonic solids, namely the tetrahedron, the cube, the octahedron, the dodecahedron and the icosahedron. 
The 5 Platonic solids are shown in  Figure $\ref{fig:Platonic}$. These solids have an underlying network structure which is planar as shown in Figure $\ref{fig:planar}$.
\item[(4)]{\bf Dimension  $d= 4$}\\
In dimension $d=4$ the number of regular polytopes is 6, namely  the pentachoron, the tesseract, the hexadecacoron, the 24-cell, the 120-cell and the 600-cell.
\item[(5)]{\bf Dimension  $d\geq 5$}\\
In dimension $d\geq 5$ the number of regular polytopes is 3, i.e. the $d$-simplex, the $d$-hypercube and the $d$-orthoplex. 
\end{itemize}

 Here in the following we introduce some structural properties of the higher order networks (simplicial complexes and cell-complexes)  that will play a key role in the following paragraphs.

Let us  assign to each $\delta$-dimensional face $\alpha$ of the pure cell-complex a {\it generalized degree}   $\kappa_{d,\delta}(\alpha)$ indicating how many $d$-dimensional regular polytopes are incident to the face.
Additionally we associate to each   $(d-1)$ face of the cell-complex an {\it incidence number $n_{\alpha}$}  equal to the  generalized degree $\kappa_{d,d-1}(\alpha)$ of the same $(d-1)$-face $\alpha$ minus one, i.e.
$$n_{\alpha}=\kappa_{d,d-1}(\alpha)-1.$$

Being a $0$-dimensional face, every node $i\in\{1,2,\ldots, N\}$ of the cell-complex is also  assigned a {\it generalized degree} $\kappa_{d,0}(i)$ indicating how may $d$-dimensional regular polytopes are incident to it. The  degree $k_i$ of node $i$ is related to the generalized degree  $\kappa_{d,0}(i)$ by 
\bea
k_i= v + (v-d+1)(\kappa_{d,0}(i)-1),
\label{kkappa}
\eea
where $v$ is the degree of each node in the regular polytope.
Finally we note that here we will focus mainly on network  of   pairwise interactions induced by   the higher order network, i.e. we will mostly focus on its {\em skeleton}. 

\begin{table}
\caption{Properties of all regular polytopes in dimension $d$: $M$ number of nodes,  $F$ number of faces, $f$ number of $(d-1)$ faces incident to a node, $m$ number of nodes incident to a single $(d-1)$ face, $v$ degree of a node}
\label{table:polytopes}
\centering
\begin{tabular}{llllll}
\hline\noalign{\smallskip}
                  &       &       &           &       &                                                     \\
                     & $F$   & $M$   & $f$       & $m$    &     $v$  \\  
                     &       &       &           &        &                                                \\   
                                       \hline\noalign{\smallskip}
                        &       &       &           &      &                                                   \\
                  
$d=1$                   &       &       &           &      &                                                       \\
link                & $1$   & $2$   & $1$       & $1$     &   $1$          \\     
   &       &       &           &       &                           \\           
                     \hline\noalign{\smallskip}
                        &       &       &           &      &                          \\
                  
$d=2$                   &       &       &           &       &                           \\
$p$-polygon                 & $p$   & $p$   & $2$       & $2$   &   $2$          \\
                   &       &       &           &     &                  \\
\hline\noalign{\smallskip}
                     &       &       &           &     &                 \\
$d=3$              &       &       &           &    &                    \\
tetrahedron      & $4$   & $4$   & $3$       & $3$   &  $3$         \\
cube             & $6$   & $8$   & $3$       & $4$   &  $3$      \\
octahedron       & $8$   & $6$   & $4$       & $3$   &  $4$   \\
dodecahedron     & $12$  & $20$  & $3$       & $5$   &  $3$      \\
icosahedron      & $20$  & $12$  & $5$       & $3$   &  $5$  \\
                   &       &       &           &      &                    \\ 
\hline\noalign{\smallskip}
                     &       &       &           &      &                     \\
$d=4$             &      &      &           &     &                  \\
 pentachoron    & $5$   & $5$   & $4$      &$4$  & $4$\\
tesseract      & $8$   & $16$  & $4$       & $8$ &  $4$     \\
hexadecachoron & $16$  & $8$   & $8$       & $4$  &  $6$       \\
24-cell        & $24$  & $24$  & $6$       & $6$   &  $8$      \\
120-cell       & $120$ & $600$ & $4$       & $20$  &   $4$       \\
600-cell       & $600$ & $120$ & $20$      & $4$   &   $12$    \\        
                   &       &       &           &         &                   \\
\hline\noalign{\smallskip}
                     &       &       &           &        &                  \\
$d>4$                 &       &       &           &       &                   \\
simplex              & $d+1$ & $d+1$ & $d$       & $d$   &  $d$    \\
cube                 & $2d$  & $2^d$ & $d$       & $2^{d-1}$ & $d$    \\
orthoplex            & $2^d$ & $2d$  & $2^{d-1}$ & $d$   &  $2(d-1)$  \\     
                   &       &       &           &         &               \\
\hline   
\end{tabular}
\end{table}
\section{Network Geometry with Flavor}

The Network Geometry with Flavor \cite{NGF,Hyperbolic} is a non-equilibrium model describing the evolution of higher order networks.
Originally this model has been formulated to study the evolution and the emergent geometry of simplicial complexes, here we extend the model to pure cell-complexes formed by identical  regular $d$-dimensional  polytopes.  

The Network Geometry with Flavor  depends on the specific regular polytope that form its building blocks and in particular on its dimension $d$. Moreover it also depends on  a parameter $s$ called {\it flavor} taking values $s\in\{-1,0,1\}$.

The algorithm  generating the Network Geometry with Flavor, is simply stated. \\
At time $t=1$ the higher-order network ${\mathcal Q}$ is formed by a single regular polytope.\\
At each time $t>1$ we choose a $(d-1)$-dimensional face $\alpha$ of the higher order network with probability
\bea
\Pi_{\alpha}^{[s]}=\frac{1}{Z^{[s]}}(1+sn_{\alpha})
\eea
with 
\bea
Z^{[s]}=\sum_{\alpha'\in {\mathcal Q}}(1+sn_{\alpha'})
\eea
and we glue a new regular polytope to it.\\
In this model, the  necessary and sufficient (combinatorial) condition to get a discrete manifold is that every $(d-1)$-face of the higher order network has incidence network $n_{\alpha}\in \{0,1\}$.
The higher network topology generated by this model depends on the flavor $s$ and on specific type of regular polytope that forms the building block of the structure.
Here we discuss the major effect of considering different flavors.
\begin{itemize}
\item[(1)]{\bf Flavor $s=-1$}\\
 In this case we  can attach a $d$-dimensional regular polytope only to a face with  $n_{\alpha}=0$. In fact for $n_{\alpha}=1$ we have $\Pi_{\alpha}^{[-1]}=0$. Therefore each face of the higher order network will have a incidence number $n_{\alpha}\in\{0,1\}$ resulting in  a discrete manifold structure. We call these networks Complex Network Manifolds \cite{CQNM}.
 \item[(2)]{\bf Flavor $s=0$}\\
In this case $\Pi_{\alpha}^{[0]}$ is constant for each face of the higher order network. Therefore the attachment probability enforces a {\it  uniform attachment} in which every face has the same probability to attract new regular polytopes. Consequently the incidence number can take any value $n_{\alpha}\in \mathbb{N}^0$.
 \item[(3)]{\bf Flavor $s=1$}\\
In this case the probability $\Pi_{\alpha}^{[1]}$  to attach a new regular polytope to the face $\alpha$ is proportional to the generalized degree of the face $\kappa_{d,d-1}(\alpha)=1+n_{\alpha}$, resulting in a {\it explicit preferential attachment mechanism}.
Consequently the incidence number can take any value $n_{\alpha}\in \mathbb{N}^{0}$.
\end{itemize}

The Network Geometry with Flavor  been proposed in Ref. \cite{NGF} and Complex Network Manifolds have been first introduced in \cite{CQNM,PRE} for describing the evolution and growth of simplicial complexes.  However the Network Geometry with Flavor reduces to other known models in some specific limits.
\begin{itemize}
\item[(1)]{\bf Dimension  $d=1$}\\
In dimension $d=1$ the Network Geometry with Flavor is a growing tree and reduces for $s=-1$ to a growing chain, for $s=0$ to a tree growing by uniform attachment, and for $s=1$ it reduces to the Barab\'asi-Albert model with preferential attachment \cite{BA}.
\item[(2)]{\bf Dimension  $d=2$}\\
 The Network Geometry with Flavor $s=0$ having triangles as building blocks  has been first proposed in Ref. \cite{Doro_triangles}.
\item[(3)]{\bf Dimension  $d=3$}\\
In dimension $d=3$  the Network Geometry with Flavor  $s=-1$  reduces to a random  Apollonian network \cite{apollonian,apollonian_ising,apollonian2,Apollonian_group}.
\end{itemize}
Therefore the Network Geometry with Flavor can be considered as a theoretical framework which unifies and extends several well known network models. Moreover as we will see in the next section it reveals an important mechanism for emergent hyperbolic network geometry.

We observe that  variations of this model can be envisaged in the following directions:
\begin{itemize}
\item[(i)]
The present choice of the  values $s\in \{-1,0,1\}$ for the flavor $s$ is driven by the need to explore  regions of the  possible parameter space with very distinct dynamics. Note however that the model can be as well studied by taking any real positive value of $s$ (which will enforce a preferential attachment with initial attractivness of the faces\cite{Doro_book}) or any rational negative value of $s$ with $-1\leq s<0$ (which will enforce a upper limit to the  number of polytopes that are incident to any given face).
\item[(ii)]
The model can be easily extended to cell-complexes that are not pure by allowing the gluing of regular polytopes having the same $(d-1)$-faces. For instance it is possible to consider a variation of the $d=3$ Network Geometry with Flavor in  which tetrahedra, octahedra and icosahedra can be glued  along their  triangular faces.
\item[(iii)]
The model can be extended by associating a  {\it fitness} to the faces of the cell-complexes and modifying the attachment probability along the lines proposed in Refs. \cite{NGF,Hyperbolic}. This modification can lead to very interesting topological phase transitions.
\end{itemize}

Despite the fact that these modifications of the model have significant potential for understanding network geometry, in this paper, due to space limitation, we limit our study to pure cell-complexes described by the  Network Geometry with Flavor  $s\in\{-1,0,1\}$ in which we do not consider the effect of the fitness of the faces.
 
\section{Emergent Hyperbolic Geometry}

While the definition of the Network Geometry with Flavor is purely topological, the emergent geometry is observed when one attributes equal length to all the links. Attributing the same length to each link consists of making the least biased assumption on their length. Therefore this procedure defines the main path to explore the emergent hidden network geometry of the Network Geometry with Flavor which is a combinatorial network model that makes no explicit use of the hidden geometry.

The Network Geometry with Flavor are  small world \cite{SW} for every flavor $s$ and any dimension $d$ except from the special case $s=-1,d=1$ in which the resulting network is a chain \cite{Ana}. Specifically in Network Geometry with Flavor both the diameter $D$ and the average shortest distance increase logarithmically with the total number of nodes $N$. This implies that the number of nodes in the network $N$ increases exponentially with its diameter $D$, i.e. $N\simeq e^{\alpha D}$ where $\alpha>0$. 
Consequently, as long as we do not allow "crossing" of the simplices,  their emergent geometry cannot be an Euclidean geometry with finite  Hausdorff dimension $d_H$ because  in this case, we would observe the power-law scaling $N\simeq D^{d_H}$. This observation implies that actually the Hausdorff dimension $d_H$ of the Network Geometry with Flavor is infinite $d_H=\infty$, with the only exception of the case $s=-1,d=1$ in which $d_H=1$.

\begin{figure}
  \includegraphics[width=0.98\columnwidth]{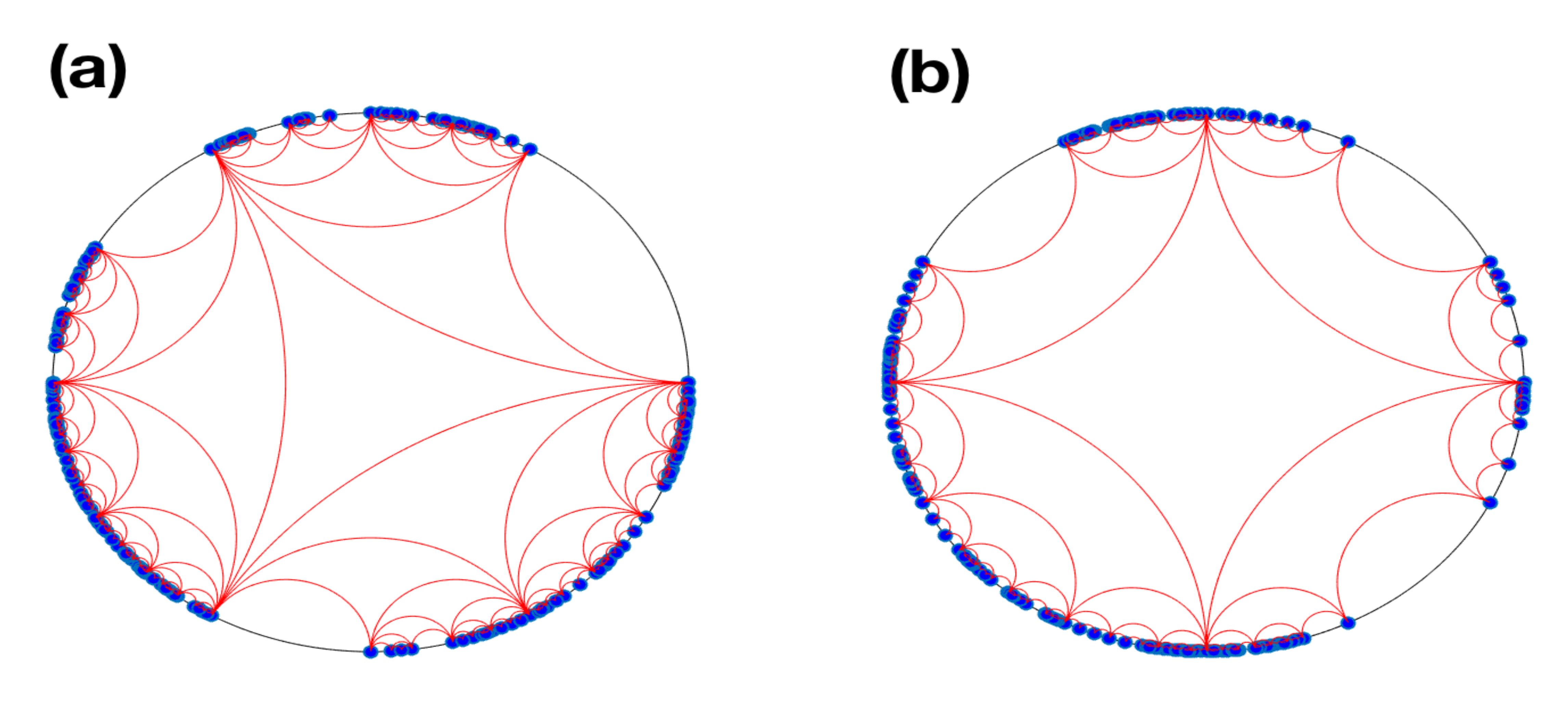}
\caption{The emergent hyperbolic geometry of $2$-dimensional Network Geometry with Flavor is shown for Network Geometry formed by triangles (panel a) and squares (panel b). The networks have flavor $s=-1$ and $N=100$ number of nodes.}
\label{fig:vis}      
\end{figure}

\begin{figure}
  \includegraphics[width=0.98\columnwidth]{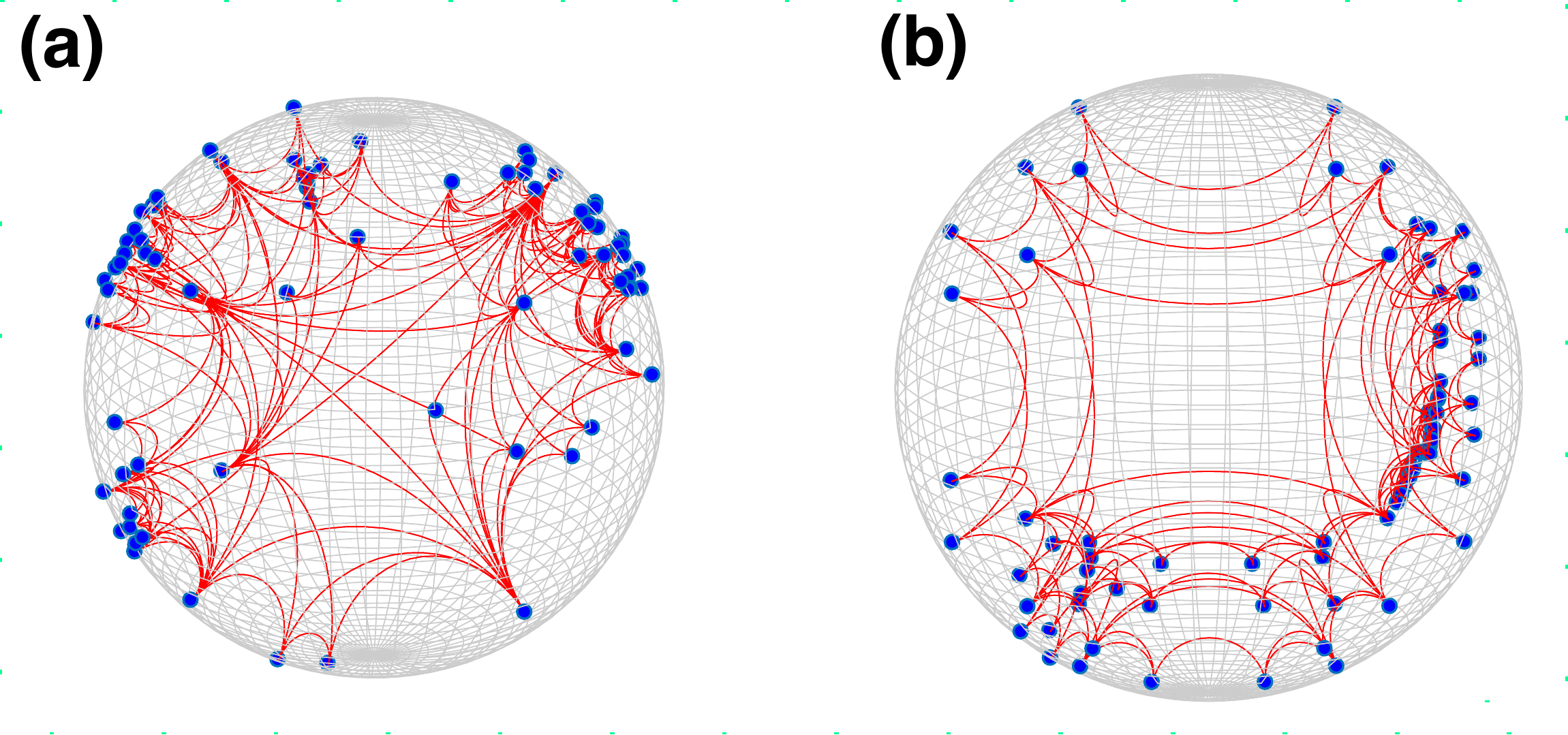}
\caption{The emergent hyperbolic geometry of $3$-dimensional Network Geometry with Flavor is shown for Network Geometry formed by tetrahedra (panel a) and cube (panel b). The networks have flavor $s=-1$ and $N=200$ number of nodes.}
\label{fig:vis2}      
\end{figure}

The emergent network geometry of the Network Geometry with Flavor is hyperbolic \cite{Hyperbolic} as long as $(s,d)\neq (-1,1)$.
This can be shown by constructing the natural hyperbolic embedding of the Network Geometry with Flavor 
in the hyperbolic spaces ${\mathbb {H}}^d$, and specifically the Poincar\'{e} ball model \cite{Hyperbolic}.   
Let us   consider a Poincar\'e ball model of $\mathbb{H}^d$. The Poincar\'e ball model includes all the points of the unit ball $B^n=\{{\bf x}\in\mathbb{R}^d: |{\bf x}|<1\}$, with $|\ldots|$ indicating the Euclidean norm. The Poicar\'e ball model is associated to the hyperbolic metric $d_B$ assigning to  each pair of points ${\bf x,y}\in \mathbb{R}^d$ the distance 
\bea
\cosh d_B({\bf x, y})=1+\frac{2|{\bf x-y}|^2}{(1-|{\bf x}|^2)(1-|{\bf y}|^2)}.
\eea
Here we identify every   $d$-dimensional polytope of our cell complex with an  ideal regular polytope of the Poincar\'e ball model. An ideal regular polytope  has all its  nodes  at the boundary of the hyperbolic ball, so all the nodes $i$ have a position ${\bf r}_i\in \mathbb{R}^d$ satisfying $|{\bf r}_i|=1$. This construction allows for having all the connected nodes at equal hyperbolic distance. Note however this distance is infinite which is the
condition we need to satisfy for having an embedding that for infinite network size fills the entire hyperbolic space.   In Figure \ref{fig:vis} and Figure \ref{fig:vis2} we show some examples of the hyperbolic embedding of Network Geometry with Flavor of dimension $d=2$ and dimension $d=3$ respectively. For dimension $d=2$ we have considered the Network Geometry with Flavor $s=-1$ formed by triangles or squares, for dimensions $d=3$ we have considered the Network Geometry with Flavor $s=-1$ formed by  tetrahedra and  cubes.


\section{Complex Network Manifolds Topological Dimensions}

The Network Geometry with Flavor has a topological dimension $d$ given by the dimension of the $d$ dimensional regular polytope that forms its  building blocks.

In particular  Complex Network Manifolds  made by $d$-dimensional  simplices are $d$-dimensional manifolds with boundary having all their nodes residing at the boundary of the manifold. Additionally the Complex Network Manifolds are $(d-1)$-connected meaning that each $d$ dimensional regular polytope can be connected to any other $d$ dimensional regular polytope by paths that go from one $d$ dimensional polytope to another one if they share a $(d-1)$-face.
Given these properties, the Complex Network Manifolds can be interpreted  as  $(d-1)$-dimensional manifolds without a boundary by considering the cell-complex formed  by all the $(d-1)$-faces with $n_{\alpha}=0$ and all their lower-dimensional faces.
In this way the  $d$-dimensional manifold can be  projected  on its $(d-1)$-dimensional boundary without losing any information about the network skeleton, i.e. while keeping all the links. 

\begin{figure}
  \includegraphics[width=0.98\columnwidth]{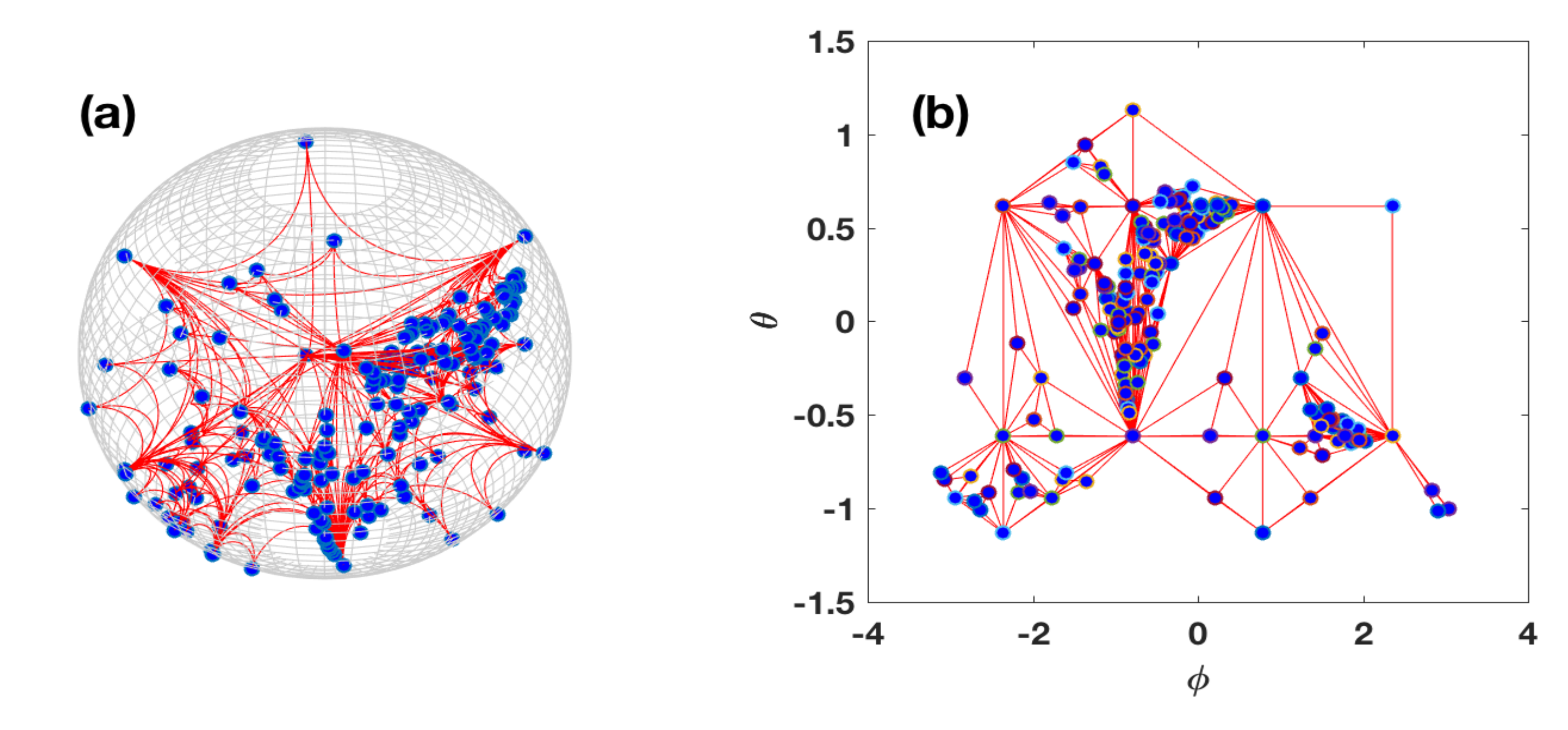}
\caption{A Complex Network Manifold in dimension $d=3$ can be studied as a $3$-dimensional manifold with boundary (panel a) or it can be projected in two dimensions in which it can be interpreted as a $2$-dimensional manifold without boundary (panel b). Here we show as an example the Complex Network Manifold formed by tetrahedra in  dimension $d=3$ and its planar projection reducing  to a random planar Apollonian network when we consider the position of the nodes in the plane $(\phi,\theta)$ corresponding to the azimuth and elevation of the Euclidean unit ball (here for clarity we have omitted to draw the links which connect nodes across the figure). Note that the Euclidean $2$-dimensional embedding of the network requires to abandon the assumption that all the links have equal length.}
\label{fig:projection}      
\end{figure}

For example Complex Network Manifolds builded by $3$-dimensional regular polytopes can be reduced to $2$-dimensional closed manifolds. Specifically  a Complex Network Manifold build from tetrahedra can be reduced to a closed $d=2$ manifold whose faces are initially four identical triangles which evolve though a sequence of successive triangulations forming a generalized Apollonian network (see  Figure $\ref{fig:projection}$).

These properties of Complex Network Manifolds reveal the "holographic character" of this model and indicate that these structures are interesting for the study of network geometry of complex networks and are closely related to tensor networks which have been attracting large interest in the quantum information community (see for instance their use in \cite{Tensor0}).

\section{Complexity and Degree Distribution}

In order to characterize the emerging complexity of the Network Geometry with Flavor, in this paragraph we derive  the degree distribution $P_{d}^{[s]}(k)$ of the Network Geometry with Flavor $s$ and dimension $d$.
In particular here our aim to to explore under which conditions on the flavor $s$, the dimensionality $d$ and the nature of the regular polytope we observe that the Network Geometry with Flavor has a scale-free topology. A scale-free network topology is  observed when  the degree distribution can be approximated for $k\gg1 $ as a power-law
\bea
P_{d}^{[s]}(k)\simeq Ck^{-\gamma}
\label{powerlaw}
\eea 
with power-law exponent
\bea
\gamma\in (2,3].
\label{SF}
\eea
This  range of power-law exponents indicates that the network is dominated by hubs nodes and the second moment of the degree distribution $\avg{k^2}$ is diverging as the network size $N$ grows, i.e. $\avg{k^2}\to \infty$ as $N\to \infty$ also if the average degree $\avg{k}$ is independent of $N$ \cite{BA,Doro_book,Newman_book,Laszlo_book}. Most notably these networks are widely represented in real complex systems and have dynamical properties strongly affected by their underlying complex scale-free topology \cite{BA,Doro_book,Newman_book,Laszlo_book}.

In order to find the degree distribution $P_d^{[s]}(k)$ let us  first derive the expression for the probability $\tilde{P}_{d}^{[s]}(\kappa)$ that a random node has generalized degree  $\kappa$ using the master equation approach \cite{Doro_book}. 

For a realization of the Network Geometry with Flavor $s$ and dimension $d$, let us indicate with  $N^{[s]}_{d}(\kappa,t)$, the average number of nodes that at time $t$ have  generalized degree $\kappa$. This quantity obeys the master equation 

\bea
    N^{[s]}_{d}(\kappa,t+1) &=& N^{[s]}_{d}(\kappa,t) + \tilde{\Pi}(\kappa-1)N^{[s]}_{d}(\kappa-1,t)(1-\delta_{\kappa,1})\nonumber \\
    &&-\tilde{\Pi}(\kappa)N^{[s]}_{d}(\kappa,t)+(M-m)\delta_{\kappa,1},
    \label{master}
\eea
where $\tilde{\Pi}(\kappa)$ is the probability that we attach a regular polytope to a $(d-1)$-face including a node with generalized degree $\kappa$, $M-m$ is the number of nodes added to the Network Geometry with Flavor at every time step and $\delta_{x,y}$ is the Kronecker delta.  

It can be shown (see Appendix for details of the derivation) that $\tilde{\Pi}(\kappa)$ is approximated for $t\gg1$ by 
\bea
\tilde{\Pi}(\kappa)=\frac{f+(f-1+s)(\kappa-1)}{(F-1+s)t}.
\label{Pikappa}
\eea
This expression indicates that in Network Geometry with Flavor,  we can observe an {\em emergent preferential attachment}. In fact for  
$f-1+s>0$ the probability $\tilde{\Pi}(\kappa)$  increases linearly with  the generalized degree $\kappa$ also when the flavor $s\in\{-1,0\}$, i.e. also when the model definition does not contain an explicit preferential attachment. Therefore in this case the emergent preferential attachment is an outcome of the network geometry.

We observe that emergent preferential attachment is  observed if and only if  the dimension $d$ satisfies $d-1+s>0$. In fact the condition $d-1+s>0$ is equivalent to the condition  $f-1+s>0$ (see Table $\ref{table:polytopes}$ for the values of $f$ as a function of $d$). Moreover we have $f-1+s=0$ only for $d-1+s=0$, i.e. only for $(d,s)=(2,-1)$  and $(d,s)=(1,0)$. Finally only for dimension $d=1$ and flavor $s=-1$ we can have $f-1+s=-1$. This case should be consider somewhat separately because the network evolution produces a one dimensional chain having only two nodes with generalized degree $\kappa=1$ and all the other nodes with generalized degree $\kappa=2$. In fact $\tilde{\Pi}(\kappa)>0$ only for $\kappa=1$.

For parameter values $(s,d)\neq (-1,1)$ the number of nodes that can be incident to the new polytope increases with the network size, generating a small-world topology. In this case we can solve the master equation using techniques extensively used for growing network models \cite{Doro_book}. 

Using Eq. $(\ref{Pikappa})$ and imposing that for parameter values $(s,d)\neq (-1,1)$ in the large network limit the number of nodes with generalized degree $\kappa$ grows as 
\bea
N^{[s]}_{d}(\kappa,t)\simeq (M-m)t\tilde{P}_{d}^{[s]}(\kappa)
\eea
we can solve the master equation (Eq. $(\ref{master})$) finding  the exact asymptotic result for the  generalized degree distribution $\tilde{P}_{d}^{[s]}(\kappa)$ valid as $t\to \infty$.  

In this way it can be shown that for 
 $d-1+s=0$  the generalized degree distribution $\tilde{P}_{d}^{[s]}(\kappa)$ is exponential  and  given by 
\bea
\tilde{P}_{d}^{[s]}(\kappa) =\frac{F-f}{F}\left(\frac{f}{F}\right)^{\kappa-1}.
    \eea
For $(d-1+s)>0$, however the generalized degree distribution $\tilde{P}_{d}^{[s]}(\kappa)$ is given by 
\bea
    \tilde{P}_{d}^{[s]}(\kappa) = \tilde{C}\frac{\Gamma\left(\kappa-1+f/(f-1+s)\right)}{\Gamma\left(\kappa+(F-1+s+f)/(f-1+s)\right)}.
\eea
where $\tilde{C}$ is a constant given by 
\bea
\tilde{C}=\frac{F-1+s}{F-1+s+f}\frac{\Gamma(1+(F-1+s+f)/(f-1+s))}{\Gamma(f/(f-1+s))}.
\eea

Using these results of the generalized degree distribution $\tilde{P}_d^{[s]}(\kappa)$ let  us now derive the degree distribution $\tilde{P}_d^{[s]}(k)$.
 
In  the case $s=-1,d=1$ the Network Geometry with Flavor is a one-dimensional chain and it is easy to see that the degree distribution is bimodal and given by 
\bea
P_d^{[s]}(1)&=&\frac{2}{N},\nonumber \\
P_d^{[s]}(2)&=&1-\frac{2}{N},
\eea
with $P_d^{[s]}(k)=0$ for every $k>2$. In fact in a chain only two nodes have degree $k=1$ and all the other nodes have degree $k=2$.

\begin{figure}
  \includegraphics[width=1.0\columnwidth]{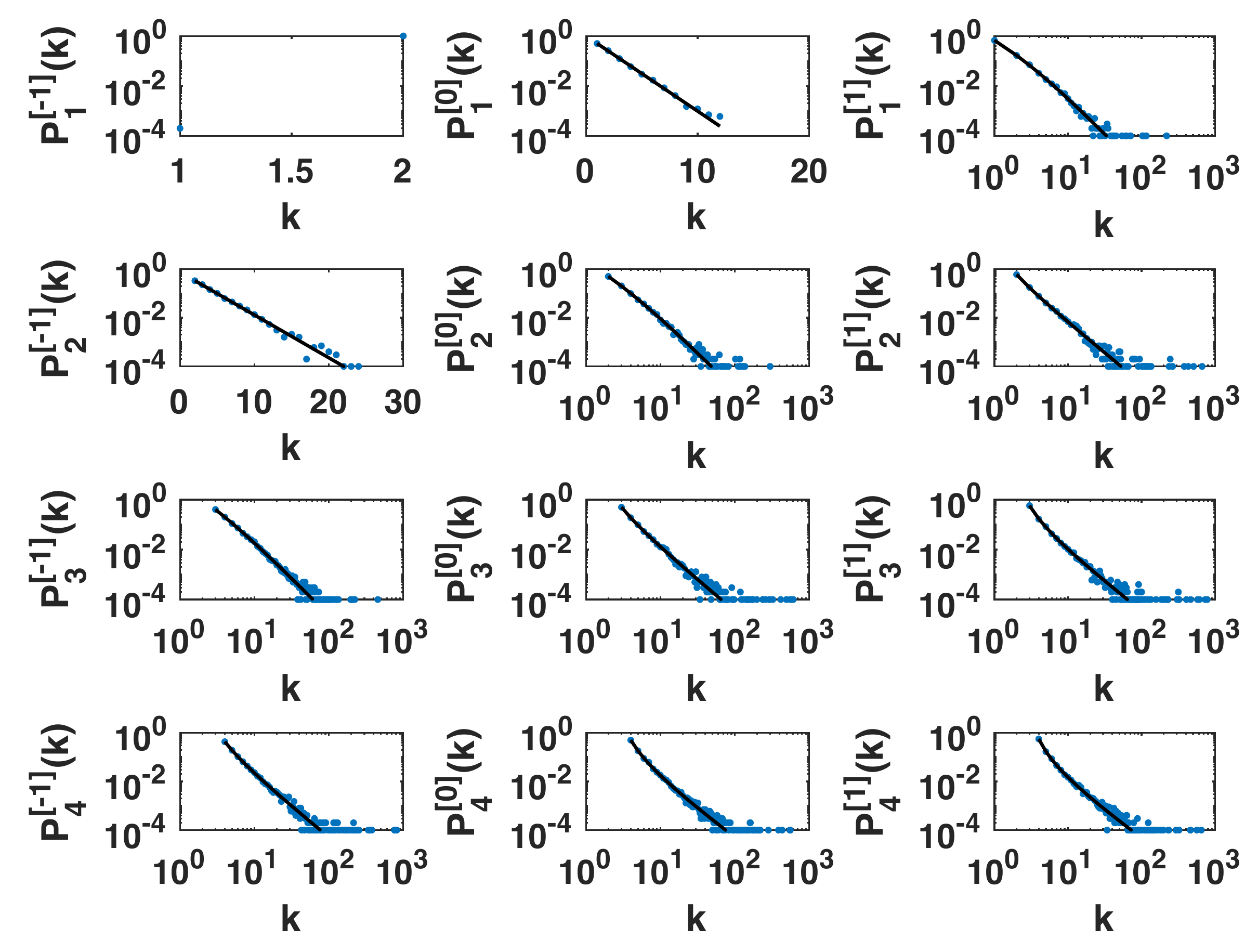}
\caption{The degree distribution $P_d^{[s]}(k)$ of the  Network Geometry with Flavor formed by simplices,  having flavor  $s=-1,0,1$ and dimension $d=1,2,3,4$ (symbols) is compared to the analytical predictions (solid line). The Network Geometry with Flavor has $N=10^4$ nodes.}
\label{fig:distribution1}      
\end{figure}

\begin{figure}
  \includegraphics[width=1.0\columnwidth]{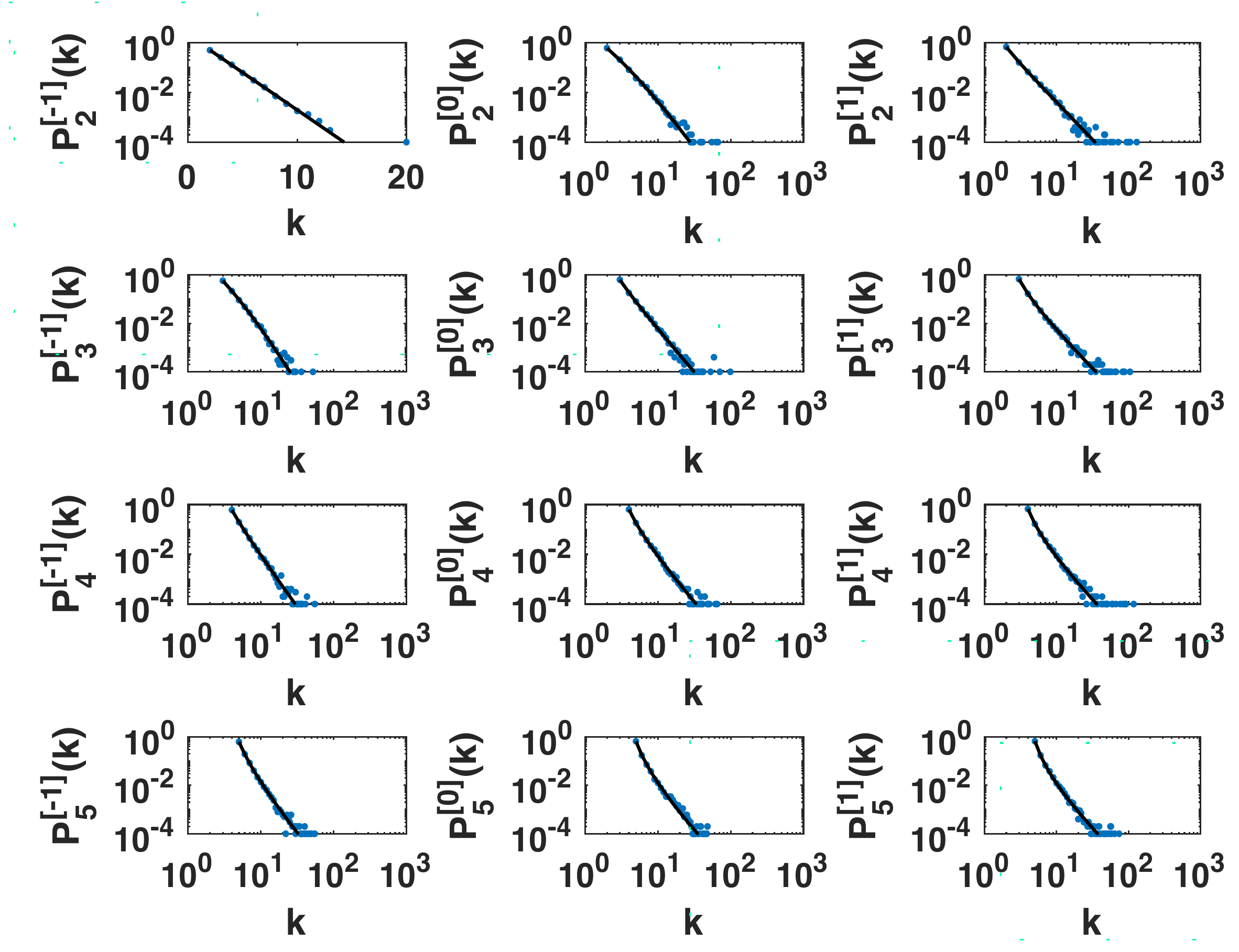}
\caption{The degree distribution $P_d^{[s]}(k)$ of the  Network Geometry with Flavor formed by hypercubes, having flavor  $s=-1,0,1$ and dimension $d=2,3,4,5$ (symbols) is compared to the analytical predictions (solid line). The Network Geometry with Flavor has $N=10^4$ nodes.}
\label{fig:distribution2}      
\end{figure}

\begin{figure}
  \includegraphics[width=1.0\columnwidth]{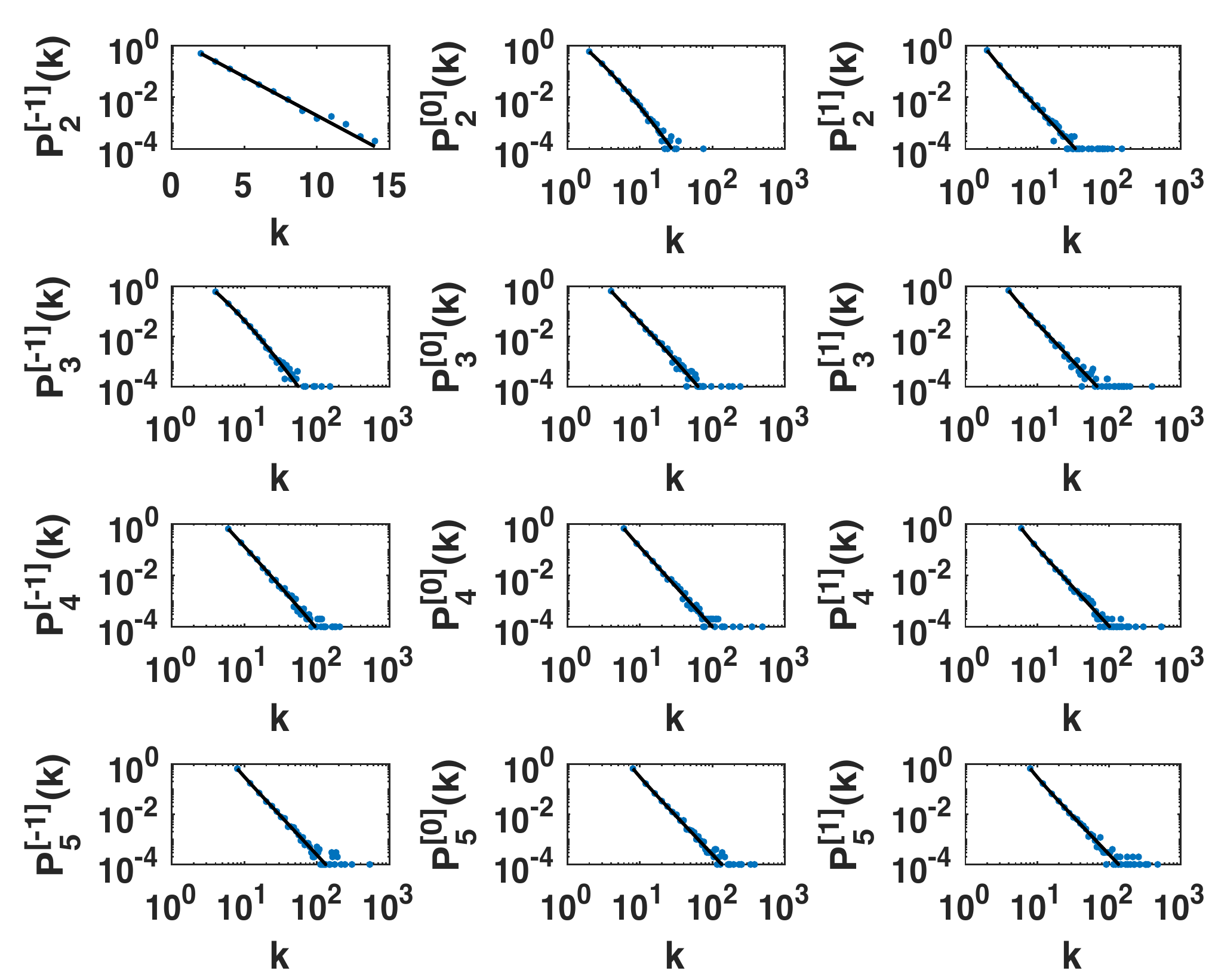}
\caption{The degree distribution $P_d^{[s]}(k)$ of the  Network Geometry with Flavor formed by orthoplexes, having flavor $s=-1,0,1$ and dimension $d=2,3,4,5$ (symbols) is compared to the analytical predictions (solid line). The Network Geometry with Flavor has $N=10^4$ nodes.}
\label{fig:distribution3}      
\end{figure}

For $(s,d)\neq(-1,1)$,  using Eq. ($\ref{kkappa}$) we can derive the expression for the degree distribution $P_d^{[s]}(k)$ in terms of the generalized degree distribution $\tilde{P}_d^{[s]}(\kappa)$
\bea
    P_{d}^{[s]}(k) &=&\tilde{P}_d^{[s]}\left(\frac{k-v}{v-d+1}+1\right).
    \eea    
    Therefore for $d-1+s=0$ (for which $f=d$) the degree distribution is exponential and given by 
    \bea
   {P}_{d}^{[s]}(k) =\frac{F-f}{F}\left(\frac{f}{F}\right)^{{k-v}},
    \eea
    while for $d-1+s>0$ it is given by    
    \bea
    \nonumber \\
    {P}_{d}^{[s]}(k)   &=&\tilde{C}\frac{\Gamma\left[({k-v})/({v-d+1})-1+f/(f-1+s)\right]}{\Gamma\left[({k-v})/({v-d+1})+(F-1+s+f)/(f-1+s)\right]}.
\eea
Therefore for $d-1+s>0$ the degree distribution decays as a power-law, i.e. follows Eq. $(\ref{powerlaw})$
with power-law exponent $\gamma$ given by 
\bea
   \gamma = 1 + \frac{F-1+s}{f-1+s}.
   \label{gamma}
\eea
In Figure $\ref{fig:distribution1}$, $\ref{fig:distribution2}$  and Figure $\ref{fig:distribution3}$ we show the  agreement between these theoretical expectations and the degree distribution of simulated Network Geometry with Flavor $s\in \{-1,0,1\}$ and dimension $d\leq 5$ built by simplices, hypercubes and orthoplexes respectively.

From this analytical derivation of the degree distribution it follows that the Network Geometry with Flavor has power power-law  degree distribution if and only if
\bea
d>1-s.
\eea
In Table \ref{table2} we summarize the functional form (Bimodal, Exponential, Power-law) of the Network Geometry with Flavor as a function of the dimension $d$ and the flavor $s$. Note that this classification is valid for any Network Geometry with Flavor having dimension $d$ and flavor $s$ independently of the specific regular polytope that forms its building blocks.
\begin{table}
\center
\label{table2}
\caption{Degree distribution of Network Geometry with Flavor as a function of the flavor $s$ and the dimension $d$. }
\footnotesize
\begin{tabular}{@{}llll}
\hline
& & & \\
flavor &$s=-1$&$s=0$&$s=1$\\
& & & \\
\hline 
& & & \\
$d=1$&Bimodal&Exponential&Power-law\\
& & & \\
\hline
& & &\\
$d=2$&Exponential&Power-law& Power-law\\
& & & \\
\hline
& & &\\
$d\geq 3$&Power-law&Power-law& Power-law\\
& & & \\
\hline
\end{tabular}
\end{table}

\begin{table}
\caption{Power-law exponent $\gamma$ of the degree distribution of Network Geometry with Flavor  $s$ built by gluing  regular, convex polytopes in dimension $d$. }
\label{tab:exponents}
\centering
\begin{tabular}{llll}
\hline\noalign{\smallskip}
                       &                                &                              &                   \\
                   $\gamma$      & $s= -1$              & $s=0$              &$s=1$ \\  
                       &                                &                              &                   \\   
                                       \hline\noalign{\smallskip}
                          &                                &                              &                   \\
                  
$d=1$                          &                                &                              &                   \\
link                     & N/A                & N/A               & $3$   \\     
             &                                &                              &                   \\           
                     \hline\noalign{\smallskip}
                                  &                                &                              &                   \\
                  
$d=2$                           &                                &                              &                   \\
$p$-polygon                     & N/A                       & $p$                          & $1+\frac{p}{2}$   \\
                        &                                &                              &                   \\
\hline\noalign{\smallskip}
                         &                                &                              &                   \\
$d=3$                          &                                &                              &                   \\
 tetrahedron            & $3$                            & $2\frac{1}{2}$               & $2\frac{1}{3}$    \\
 cube                   & $5$                            & $3\frac{1}{2}$               & $3$               \\
 octahedron            & $4$                            & $3\frac{1}{3}$               & $3$               \\
 dodecahedron         & $11$                           & $6\frac{1}{2}$               & $5$               \\
 icosahedron            & $7$                            & $5\frac{3}{4}$               & $5$               \\
                            &                                &                              &                   \\ 
\hline\noalign{\smallskip}
                          &                                &                              &                   \\
$d=4$                      &                                &                              &                   \\
 pentachoron          & $2\frac{1}{2}$                 & $2\frac{1}{3}$               & $2\frac{1}{4}$    \\
 tesseract           & $4$                            & $3\frac{1}{3}$               & $3$               \\
 hexadecachoron      & $3\frac{1}{3}$                 & $3\frac{1}{7}$               & $3$               \\
 24-cell             & $6\frac{1}{2}$                 & $5\frac{3}{5}$               & $5$               \\
 120-cell         & $60$                           & $40\frac{2}{3}$              & $31$              \\
 600-cell            & $34\frac{2}{9}$                & $32\frac{10}{19}$            & $31$      \\        
                           &                                &                              &                   \\
\hline\noalign{\smallskip}
                              &                                &                              &                   \\
$d>4$                           &                                &                              &                   \\
simplex                    & $2+\frac{1}{d-2}$            & $2+\frac{1}{d-1}$            & $2+\frac{1}{d}$ \\
cube                  & $3+\frac{2}{d-2}$         & $3+\frac{1}{d-1}$         & $3$               \\
orthoplex              & $3+\frac{1}{2^{(d-2)}-1}$ & $3+\frac{1}{2^{d-1}-1}$ & $3$     \\     
                        &                                &                              &                   \\
\hline   
\end{tabular}
\end{table}
If  we make the distinction between scale-free degree distributions with power-law exponents $\gamma\in (2,3]$ and more homogeneous power-law exponents $\gamma>3$ we notice that not only the dimensionality of the regular polytope but also its geometry has important consequences.

In the case of simplicial complexes,  the power-law degree distributions of the Network Geometry with Flavor are always scale-free.
This implies   that explicit preferential attachment imposed by the flavor $s=1$ always gives rise to scale-free simplicial complexes topologies with power-law exponent $\gamma\in (2,3]$. Moreover this result  indicates that  the observed emergent preferential attachment occurring for $s\in\{0,1\}$ implies that both  simplicial Complex Network Manifolds (flavor $s=-1$) and simplicial complexes evolving by  uniform attachment (flavor $s=0$) are scale-free, provided that  the dimension is sufficiently high. In fact the emergent preferential attachment is observed only for $d-1+s>0$.

However when we include the treatment of Network Geometry with Flavor formed by any type of regular polytope the rich interplay between network geometry and complexity is revealed and a much more nuanced scenario emerges.

The expression for the power-law exponent $\gamma$ (Eq. $(\ref{gamma})$) together with the condition to get a scale-free distribution (Eq. $(\ref{SF})$) indicates that the Network Geometry with Flavor $s$ and dimension $d>1-s$ are scale-free  only if 
\bea
\frac{F-1+s}{f-1+s}\leq 2.
\eea
This relation implies the following dependence of the scale-free property with the dimension $d$ and the flavor $s$.
\begin{itemize}
\item[(1)]{\bf Flavor   $s=-1$}\\
In dimension $d\geq 3$ the Network Geometry with Flavor are power-law distributed. However only the simplicial complexes are scale-free.
\item[(1)]{\bf Flavor   $s=0$}\\
In dimension $d\geq 2$ the Network Geometry with Flavor are power-law distributed. However only the simplicial complexes are scale-free.
\item[(3)]{\bf Flavor   $s=1$}\\
The Network Geometry with Flavor are always power-law distributed. For dimension $d=1$ and $d\geq 4$ Network Geometry with Flavor  $s=1$ implying an explicit preferential attachment are always scale-free. However for dimension $d\in \{2,3,4\}$ they  are not always  scale-free.\\
\begin{itemize}
\item For $d=2$  the  Network Geometry with Flavor $s=1$  are not scale-free   if they are formed by polygons different from triangles and squares. \\
\item For $d=3$  the  Network Geometry with Flavor $s=1$ are not scale-free if they are formed by dodecahedra and icosahedra.\\
\item For $d=4$ the   Network Geometry with Flavor $s=1$ are not scale-free if they are formed by  the 24-cell, the 120-cell and the 600-cell.
\end{itemize}
\end{itemize}

\section{Complexity and Emergent Community Structure}

\begin{figure}
  \includegraphics[width=0.98\columnwidth]{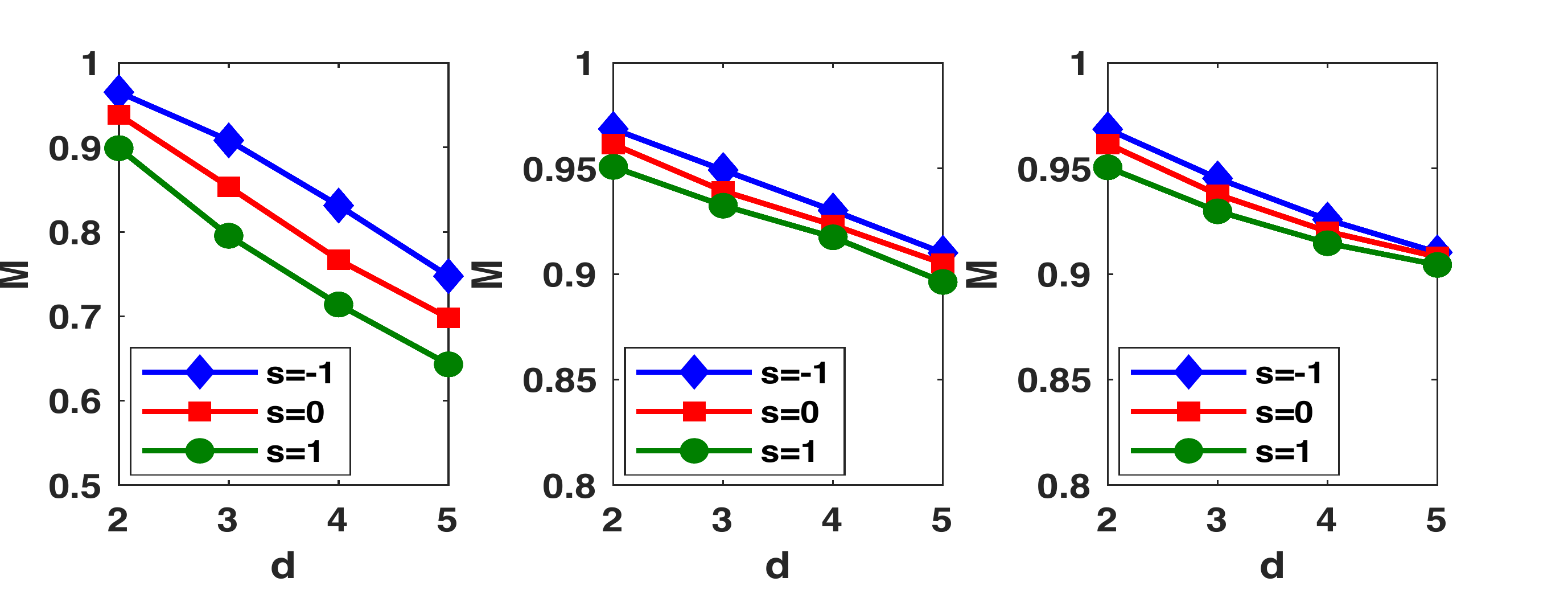}
\caption{The emergent community structure of Network Geometry with Flavor is revealed by calculating the average modularity $M$ found by the GenLouvain algorithm \cite{Gen_Louvain,Louvain} as a function of the dimension $d$ for different flavors $s=-1,0,1$. The modularity is averaged over 10 realizations of the Network Geometry with Flavor with $N=10^4$ nodes. Panel (a) refers to the Network Geometry with Flavor constructed from $d$-dimensional simplices, panel (b) refers to the Network Geometry with Flavor constructed from $d$-dimensional hypercubes and panel (c) to the Network Geometry with Flavor constructed from $d$-dimensional orthoplexes. }
\label{fig:modularity}      
\end{figure}

An important  signature  of the complexity of the Network Geometry with Flavor is its emergent community structure. In fact this model, constraining the microscopic structure of the network formed by identical, highly clusterised building blocks (the regular polytopes), spontaneously generates a mesoscale structure organized in  communities of nodes more densely connected with each other than with the other nodes of the network \cite{Iacovacci,Hyperbolic}. 
In order to characterize the emergent mesoscale structure of the Network Geometry with Flavor we have estimated the maximal  modularity $M$ \cite{Newman_book} of the network by averaging    the  results obtained using   the  GenLouvain algorithm \cite{Gen_Louvain,Louvain}  over different realizations of the   Network Geometry with Flavor having up to dimension $d=5$  (see Figure \ref{fig:modularity}).
From Figure \ref{fig:modularity} it is possible to appreciate  that while the  modularity $M$  decreases as the  topological dimensions $d$ increases, its values remain significant for every flavor $s$ up to dimension $d=5$.

A  non-trivial community structure  is observed very widely in network data.
Therefore the emergent community structure of the  Network Geometry with Flavor is a desired property for the modelling of real complex networks   observed also in other growing network models \cite{Iacovacci,Hyperbolic,Redner,Bagrow}.
 However  the community structures of real datasets  can display significant differences for different networks. Therefore here,  it is not our intention to fit the Network Geometry with Flavor to any specific real data, rather our aim is  to indicate  that the Network Geometry with Flavor  can provide a simple stylized mechanism to generate a discrete network structure with communities.

\section{Spectral  Dimension of Network Geometry with Flavor}

The spectral dimension $d_S$ \cite{Burioni,Raffaella2,Spectral,Benedetti,Thomas} of a networks characterizes how the structure of the network and its underlying network geometry affects the property of diffusion and has profound implications for quantum networks as well \cite{Piilo}. 
The  Laplacian matrix ${\bf L}$ of the network of elements 
\bea
L_{ij}=\delta_{ij}k_i-{a_{ij}},
\eea 
where ${\bf a}$ indicates the adjacency matrix of the network characterizes fully the properties of diffusion on a given network.
In fact the probability diffusion of a given continuous variable $x_i$ defined on each node $i$  of the network follows 
\bea
\frac{dx_i(t)}{dt}=-\sum_{j=1}^N L_{ij}x_j(t).
\eea
with given initial condition $x_i(0)=\delta_{i,i_0}$ describing the initial concentration of the continuous variable ${\bf x}$ on the node $i_0$.
The spectral properties of the  Laplacian fully determine the diffusion properties.
The  Laplacian  has real spectrum with eigenvalues $0=\lambda_1\leq \lambda_2\ldots \leq \lambda_N$. The degeneracy of the zero eigenvalue is equal to the number of connected components of the network. Therefore for  Network Geometry with Flavor the zero eigenvalue is not degenerate and $\lambda_2>0$.
Let us indicate with  $\rho(\lambda)$ the density of eigenvalues. The spectral dimension, if it exist, characterizes the power-law scaling of $\rho(\lambda)$ as a function of $\lambda$
\bea
\rho(\lambda)\propto \lambda^{\frac{d_S}{2}-1},
\label{rhodef}
\eea
valid for $0<\lambda\ll1$. In particular network models with finite spectral dimension must have $\lambda_2\to 0$ as $N\to \infty$.\\
In discrete lattices the spectral dimension $d_S$ is known to be equal to the Hausdorff dimension $d_H$ of the lattice and the dimension $d$ of the unitary cell of the lattice, however the spectral dimension of a network in general is not equal to its Hausdorff dimensional and satisfies $d_S\leq d_H$ \cite{Burioni}. Additionally if we consider the skeleton of a $d$-dimensional simplicial complex in general we will not find that the spectral dimension $d_S$ is equal to $d$.

Note that not every network has a spectral dimension. Most notably in networks in which the smallest non-zero  eigenvalue $\lambda_2$ is well separated from the smallest eigenvalue $\lambda_1=0$, the spectral dimension is not defined and we say in that case that the network has a spectral gap (technically a model having a spectral gap means that $\lambda_2$ is not vanishing in the large network limit).
However the presence of a spectral dimension is the rule in  networks with a non-trivial underlying geometry like lattices and fractal structures \cite{Burioni,Raffaella2,Spectral,Benedetti,Thomas}.
While in presence of the spectral gap,  convergence to  the steady state of the diffusion dynamics is exponentially fast with a typical time scale $\tau=1/\lambda_2$, in absence of a spectral gap it can be much slower. In fact when the spectral gap closes and the network has a  finite spectral dimension  $d_S$ the density distribution at the starting node asymptotically in time decays as 
\bea
\pi(t)\propto t^{-\frac{d_S}{2}}.
\eea

\begin{figure}
  \includegraphics[width=0.98\columnwidth]{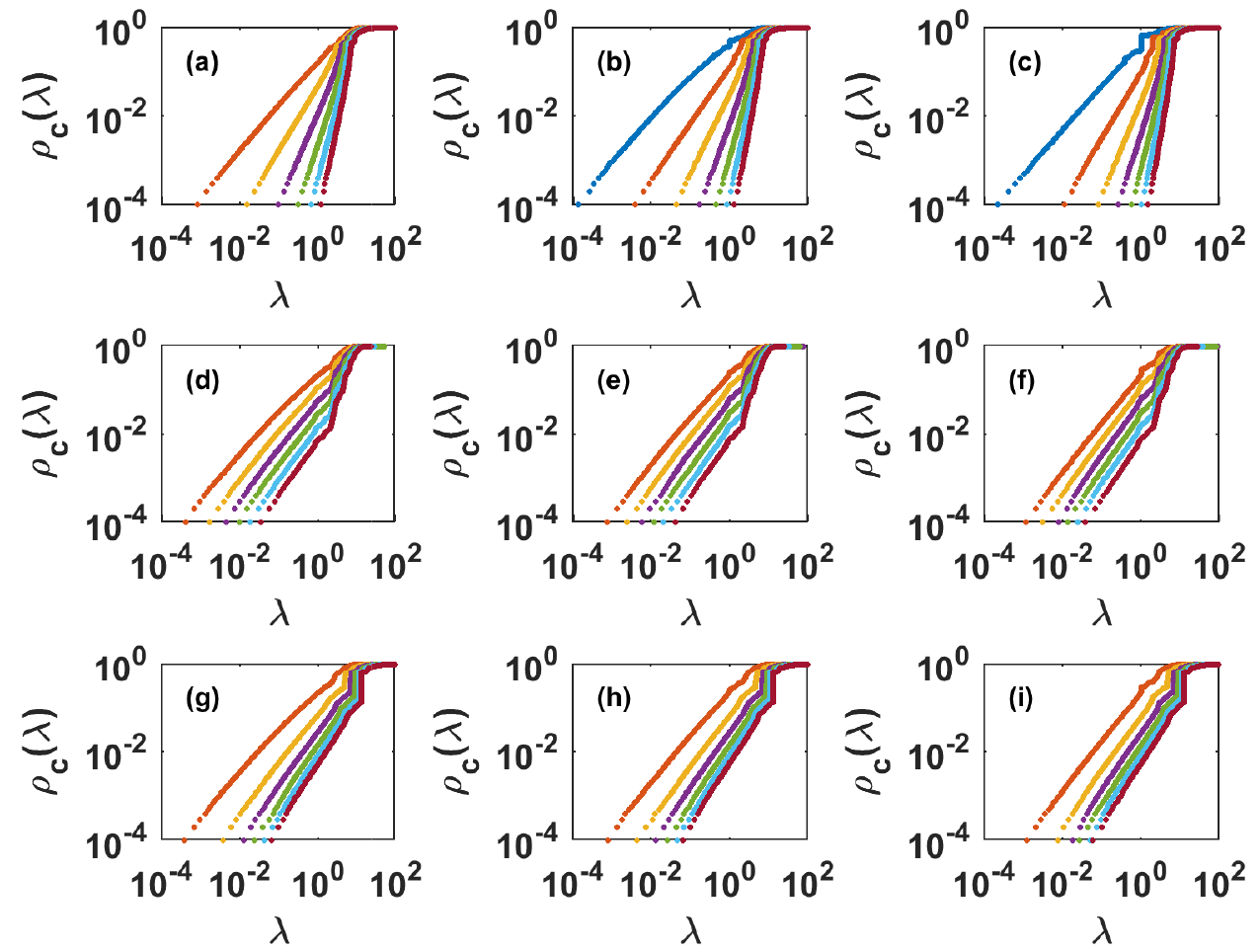}
\caption{The cumulative distribution of eigenvalues $\rho_c(\lambda)$ of the Laplacian matrix  is shown versus $\lambda$ for the Network Geometry with Flavor formed by polytopes of  dimensions  $d=1$ (blue symbols), $d=2$ (red symbols), $d=3$ (yellow symbols), $d=4$ (violet symbols), $d=5$ (green symbols), $d=6$ (light blue symbols) and $d=7$ (dark red symbols).  Panels (a), (b), (c)  refer to Complex Network Geometry formed by $d$-dimensional simplices, with flavor  $s=-1$, $s=0$ and $s=1$ respectively. Panels (d), (e), (f)  refer to Complex Network Geometry formed by $d$-dimensional hypercubes of flavor  $s=-1$, $s=0$ and $s=1$ respectively. Panels (g), (h), (i)  refer to Complex Network Geometry formed by $d$-dimensional orthoplexes of flavor   $s=-1$, $s=0$ and $s=1$ respectively. Here Network Geometry with Flavor have $N=10^4$ nodes and the cumulative distribution of eigenvalues  is averaged over $100$ network realizations. The Complex Network Manifold of  dimension  $d=1$  is only defined for simplicial complexes, and only shown for flavors $s\in\{0,1\}$ because the case $s=-1$ is  trivial (linear chain).}
\label{fig:spectral}      
\end{figure}

In presence of a spectral dimension, given that  Eq. ($\ref{rhodef}$) holds, the cumulative density of eigenvalues of the Laplacian $\rho_c(\lambda)$ obeys the scaling relation
\bea
\rho_c(\lambda)\propto \lambda^{\frac{d_S}{2}},
\eea
for $0<\lambda\ll1$.
Figure $\ref{fig:spectral}$ shows  $\rho_c(\lambda)$ for Network Geometry with Flavor $s\in\{-1,0,1\}$ formed  by $d$-dimensional simplices, $d$-dimensional hypercubes and $d$-dimensional orthoplexes up to $d=7$. From this figure it is apparent that these cell-complexes have a finite spectral dimension.

\begin{figure}
  \includegraphics[width=0.98\columnwidth]{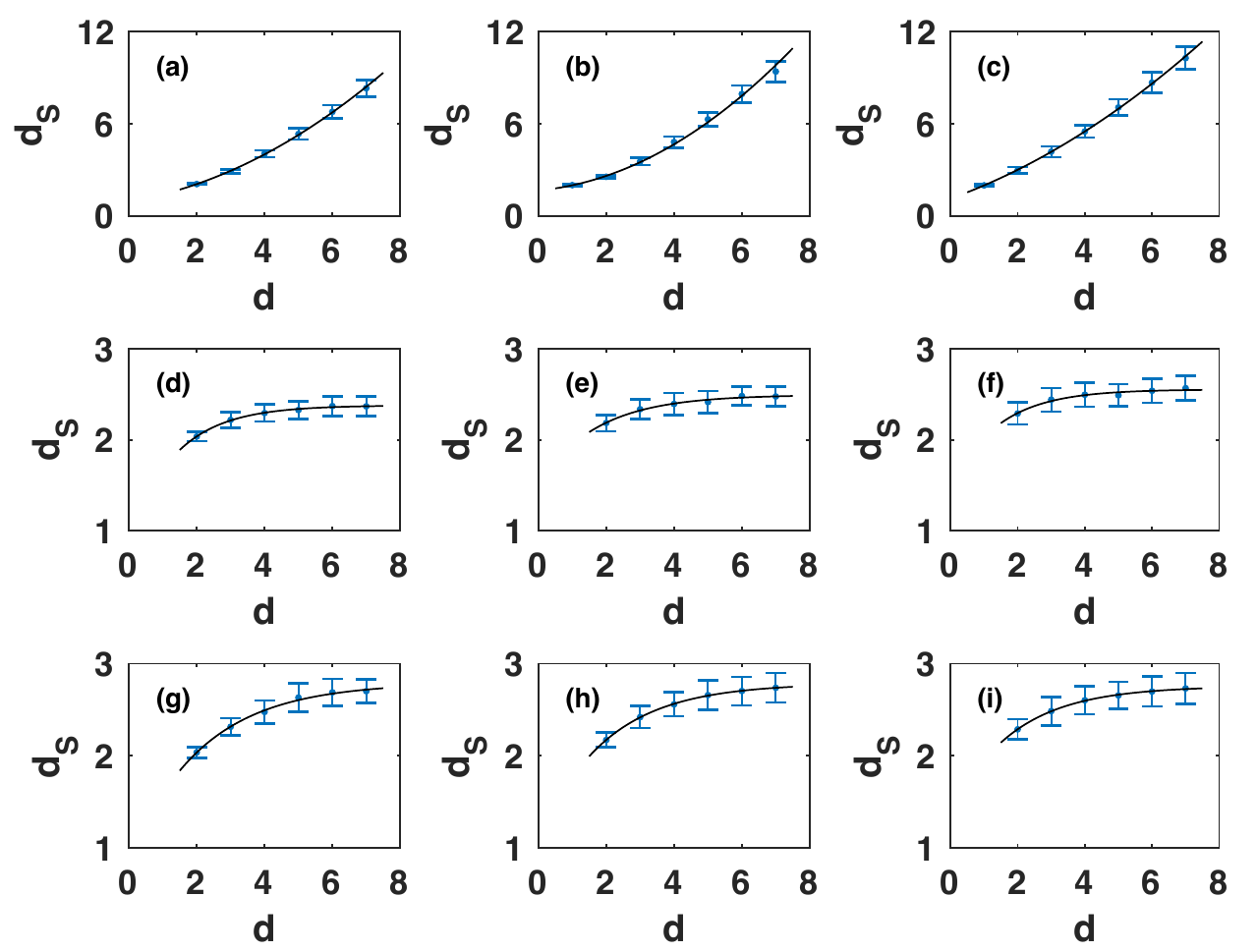}
\caption{The spectral dimension $d_S$ is shown as a function  of the topological dimension $d$ for the Network Geometry with Flavor  formed by different types of regular polytopes. Panels (a),  (b), (c)  refer to Complex Network Geometry formed by $d$-dimensional simplices, with flavor  $s=-1$, $s=0$ and $s=1$ respectively. Panels (d), (e), (f)  refer to Complex Network Geometry formed by $d$-dimensional hypercubes of flavor  $s=-1$, $s=0$ and $s=1$ respectively. Panels (g), (h), (i)  refer to Complex Network Geometry formed by $d$-dimensional orthoplexes of flavor   $s=-1$, $s=0$ and $s=1$ respectively. Here Network Geometry with Flavor have $N=10^4$ nodes and the spectral dimension $d_s$  is averaged over $100$ network realizations, the error bars indicating the standard deviation.}
\label{fig:spectralfit}      
\end{figure}
Starting from $\rho_c(\lambda)$ we have numerically evaluated the spectral dimension $d_S$ of the Network Geometry with Flavor (see Figure \ref{fig:spectralfit}) finding  that the spectral dimension $d_S$ of the Network Geometry with Flavor can be larger or smaller than $d$ depending on the value of the flavor $s$ and the nature of the polytopes that form its building blocks.
Moreover our result indicate that while for simplicial complexes the spectral dimension $d_S$ increases faster than linearly with $d$, for cell-complexes formed by hypercubes or orthoplexes the spectral dimension tends to saturate.
Specifically for simplicial complexes the spectral dimension $d_S$  can be well fitted by 
\bea
d_S = \bar{a}d^2+\bar{b}d+\bar{c},
\label{eq:quadratic}
\eea
with coefficients $\bar{a},\bar{b},\bar{c}$ depending on the flavor $s$ as shown in  Table \ref{table:fitcoeff}. Note that we have  compared the quadratic fit of $d_S$ versus $d$ to a simpler linear fit, performing a F-test, which yields very small $p$-values ($p < 0.01$) for all values of $s$, confirming the validity of the quadratic fit.
For cell-complexes formed by  $d$-dimensional hypercubes or $d$-dimensional orthoplexes, the spectral dimension $d_S$   can be fitted by 
\bea
d_S = \tilde{a}-\tilde{b}e^{-d/\tilde{c}},
\label{eq:saturation}
\eea
with coefficients $\tilde{a},\tilde{b},\tilde{c}$ shown in  Table \ref{table:fitcoeff}.

\begin{table}
\center
\label{table:fitcoeff}
\caption{Fitted coefficients of Equation (\ref{eq:quadratic}) and (\ref{eq:saturation}) for Network Geometry with Flavor formed by simplicies, hypercubes and orthoplexes. }
\footnotesize
\begin{tabular}{@{}lllll}
\hline
& & & & \\
 & &$\bar{a}$&$\bar{b}$&$\bar{c}$\\
& & & &\\
\hline
& & & & \\
Simplices& $s=-1$&0.09(1)&0.4(1)&0.8(1)\\
& & & &\\

& & & &\\
& $s=0$&0.11(2)&0.3(1)& 1.5(2)\\
& & & &\\

& & & &\\
& $s=1$&0.07(1)&0.8(1)& 1.0(1)\\
& & & &\\
\hline
& & & & \\
 & &$\tilde{a}$&$\tilde{b}$&$\tilde{c}$\\
& & & &\\
\hline
& & & & \\
Hypercubes& $s=-1$&2.38(1)&1.4(2)&1.4(2)\\
& & & &\\

& & & &\\
& $s=0$&2.49(3)&1.0(2)& 1.7(4)\\
& & & &\\

& & & &\\
& $s=1$&2.55(3)&1.0(4)& 2(1)\\
& & & &\\

& & & &\\
Orthoplexes& $s=-1$&2.79(4)&1.9(2)&0.4(1)\\
& & & &\\

& & & &\\
&$s=0$&2.78(1)&1.7(1)& 0.51(2)\\
& & & &\\

& & & &\\
& $s=1$&2.76(1)&1.4(1)& 0.54(3)\\
& & & &\\
\hline
\end{tabular}
\end{table}

These results  point out  the important role of the  regular polytope forming the building blocks of the Network Geometry with Flavor  in determining its geometrical properties.

\section{Conclusions}
In this paper we have characterized the Network Geometry with Flavor $s\in\{-1,0,1\}$ which are cell complexes built by gluing identical regular polytopes along their faces. The flavor $s=-1$ imposes that the cell complexes generated by the Network Geometry with Flavor $s=-1$ are manifolds  also called Complex Network Manifolds. The flavor $s=0$ indicates that the cell complexes grow by uniform attachment of the new polytope to a random $(d-1)$-face. The flavor $s=1$ indicates that the model includes an explicit preferential attachment of the new polytopes to $(d-1)$-faces that have large number of polytopes already attached to them.

This purely topological model generates cell complexes with emergent hyperbolic network geometry  revealed by imposing that every link has equal length.
Here we characterize the interplay between the emergent geometry of Network Geometry with Flavor and complexity.
Specifically  we characterize under which conditions the Network Geometry with Flavor are scale-free. We observe that  Network Geometry with Flavor can display or not display a scale-free degree distribution depending on the dimension $d$ flavor $s$ and specific type of regular polytope that forms its building blocks.
Interestingly the Network Geometry with Flavor which is made by simplices (and are therefore simplicial  complexes)  has notable properties that makes it different from 
other realizations of the Network Geometry formed by other types of regular polytopes.
In fact in dimension $d>2$ the simplicial complexes are scale-free for every flavor $s\in\{-1,0,1\}$ while for Network Geometry formed by other types of regular polytopes not even in presence of an explicit preferential attachment (flavor $s=1$) we are always guaranteed to obtain a scale-free degree distribution. Additionally Network Geometry with Flavor displays another important signature of complexity, i.e. they have a non-trivial emergent community structure.

Interestingly the special role of simplicial complexes is also revealed by the spectral properties of the Network Geometry with Flavor which depend on the nature of the specific regular polytope that forms its building block. For instance if the building block is a $d$-simplex we have found that the spectral dimension $d_S$ increases with the  dimension $d$, while if the building block is an $d$-dimensional hypercube and for the $d$-dimensional orthoplex the spectral dimensions tend to saturate as the dimension $d$ increases.

This work can be extended in different directions. First of all there are very clear paths leading to possible generalizations of the model  including other values of the flavor, the introduction of a fitness of the faces of the polytope or the  extention of  the model beyond pure cell-complexes. Secondly this theoretical framework provides an ideal setting to study the interplay between network geometry and dynamics  such as  frustrated synchronization \cite{Ana}. Finally this framework is very promising for  establishing close  connections between growing network models and tensor networks.
\appendix
\section*{Appendix: Derivation of Eq. ($\ref{Pikappa}$)}

In this appendix our goal is to derive Eq. ($\ref{Pikappa}$) providing the expression for the probability $\tilde{\Pi}(\kappa)$ to glue a new regular polytope which increases the generalized degree of a  node having  generalized degree $\kappa$.
Since in the Network Geometry with Flavor one polytope is added at each time step, the probability $\tilde{\Pi}_i$  that we glue  a new regular polytope to a $(d-1)$-face  $\alpha$ incident to a node $i$ (i.e. $i\subset \alpha$)  is given by 
\bea
\tilde{\Pi}_i=\sum_{\alpha|i\subset \alpha}\Pi_{\alpha}^{[s]}=\sum_{\alpha|i\subset \alpha} \frac{1+sn_{\alpha}}{Z^{[s]}}.
\label{Pii}
\eea 
For $t\gg1$ we note that  we can approximate $Z^{[s]}$ is given by 
\bea
Z^{[s]}=\sum_{\alpha \in {\mathcal Q}}(1+sn_{\alpha})=F+(F-1+s)(t-1)\simeq (F-1+s)t,
\label{Zs2}
\eea
where the last expression is the approximate expression for $t\gg1$.
 In fact  for $s=-1$  each new regular polytope introduces  $F-1$ new $(d-1)$-faces  each one  contributing one to $Z^{[s]}$. However the $(d-1)$-face $\alpha$ to which we attach the new polytope  acquires incidence number $n_{\alpha}=1$ and therefore  its contribution should be removed from $Z^{[s]}$. Therefore $Z^{[-1]}\simeq (F-2)t$. For $s=0$,  $Z^{[0]}$ simply counts the total number of different faces, therefore each new polytope contributes by a term given by $(F-1)$ to the sum corresponding the the number of novel $(d-1)$ faces that each regular polytope introduces. Therefore  $Z^{[-1]}\simeq (F-1)t$. Finally  for $s=1$, each face is counted proportional to the number of polytopes that are incident to it. Therefore any new regular polytope contributes by a term given by  $F$ and $Z^{[-1]}\simeq Ft$. \\
If a node has generalized degree $\kappa_{d,0}(i)=\kappa=1$ then it must be incident to $f$ faces $\alpha$ each one belonging to the same regular polytope. It follows that in this case the numerator of the left hand side of Eq. (\ref{Pii}) reads
\bea
\sum_{\alpha|i\subset \alpha}(1+sn_{\alpha})=f.
\eea
For  nodes with generalized degree $\kappa_{d,0}(i)=\kappa>1$, by following the same line of arguments presented above for deriving Eq. (\ref{Zs2}) it can be easily shown that 
\bea
\sum_{\alpha|i\subset \alpha}(1+sn_{\alpha})=f+(f-1+s)(\kappa-1).
\eea
 In fact each new regular polytope attached to the node after the initial one introduces $f-1$ new $(d-1)$-faces incident to node $i$ and therefore contributes by a term $(f-1+s)$ to the sum. 
 
Therefore  the probability $\tilde{\Pi}(\kappa)$ that we add a new polytope to a node with generalized degree $\kappa_i=\kappa$ is given for $t\gg 1$ by
\bea
\tilde{\Pi}(\kappa)= \frac{f+(f-1+s)(\kappa-1)}{(F-1+s)t}.
\eea




\end{document}